\documentclass[aps,prb,10pt,twocolumn,showpacs,showkeys,footinbib,superscriptaddress]{revtex4-2}

\usepackage[utf8x]{inputenc}
\usepackage{amsmath}
\usepackage{amssymb}
\usepackage{graphicx}
\usepackage[caption=false]{subfig}
\usepackage{pstricks}
\usepackage{color}
\usepackage{relsize}
\usepackage{bm}
\usepackage{braket}
\usepackage{epstopdf}

\usepackage{hyperref}
\usepackage[all]{hypcap}

\renewcommand{\i}{\ensuremath{\mathrm{i}}}
\newcommand{\e}{\ensuremath{\mathrm{e}}}
\renewcommand{\d}{\ensuremath{\mathrm{d}}}

\begin{document}

\title{Thermodynamics in topological Josephson junctions}

\author{Benedikt Scharf}
\affiliation{Institute for Theoretical Physics and Astrophysics and W\"{u}rzburg-Dresden Cluster of Excellence ct.qmat, University of W\"{u}rzburg, Am Hubland, 97074 W\"{u}rzburg, Germany}
\author{Alessandro Braggio}
\affiliation{NEST, Istituto Nanoscienze-CNR and Scuola Normale Superiore, I-56127 Pisa, Italy}
\author{Elia Strambini}
\affiliation{NEST, Istituto Nanoscienze-CNR and Scuola Normale Superiore, I-56127 Pisa, Italy}
\author{Francesco Giazotto}
\affiliation{NEST, Istituto Nanoscienze-CNR and Scuola Normale Superiore, I-56127 Pisa, Italy}
\author{Ewelina M. Hankiewicz}
\affiliation{Institute for Theoretical Physics and Astrophysics and W\"{u}rzburg-Dresden Cluster of Excellence ct.qmat, University of W\"{u}rzburg, Am Hubland, 97074 W\"{u}rzburg, Germany}

\date{\today}

\begin{abstract}
We study the thermodynamic properties of topological Josephson junctions using a quantum spin Hall (QSH) insulator-based junction as an example. In particular, we propose that phase-dependent measurements of the heat capacity offer an alternative to Josephson-current measurements to demonstrate key topological features. Even in an equilibrium situation, where the fermion parity is not conserved, the heat capacity exhibits a pronounced double peak in its phase dependence as a signature of the protected zero-energy crossing in the Andreev spectrum. This double-peak feature is robust against changes of the tunneling barrier and thus allows one to distinguish between topological and trivial junctions. At short time scales fermion parity is conserved and the heat capacity is $4\pi$-periodic in the superconducting phase difference. We propose a dispersive setup coupling the Josephson junction to a tank LC circuit to measure the heat capacity of the QSH-based Josephson junction sufficiently fast to detect the $4\pi$-periodicity. Although explicitly calculated for a short QSH-based Josephson junction, our results are also applicable to long as well as nanowire-based topological Josephson junctions.
\end{abstract}


\maketitle

\section{Introduction}\label{Sec:Intro}
Topological superconductors~\cite{Hasan2010:RMP,Qi2011:RMP,Alicea2012:RPP,Leijnse2012:SST,Tanaka2012:JPSJ,Beenakker2013:ARCMP,Tkachov2013:PSS,Culcer2020:2DM,Aguado2017:RNC,Aguado2020:PT} offer the prospect of encoding and manipulating quantum information in a fault-tolerant, topologically protected manner~\cite{Nayak2008:RMP,Kitaev2003:AP,Alicea2011:NP}. They as well as odd-frequency pairing~\cite{BlackSchaffer2012:PRB,Crepin2015:PRB,Fleckenstein2018:PRB2} and Majorana zero modes~\cite{Kitaev2001:PhysUs,Keidel2018:PRB,Fleckenstein2018:EPJ,Fleckenstein2020:arxiv} that can emerge in topological superconductors have therefore been the focus of intense research during the past decade. Proposals to realize topological superconductivity often combine $s$-wave superconductivity, magnetism, and materials with strong spin-orbit coupling~\cite{Fu2008:PRL,Lutchyn2010:PRL,Oreg2010:PRL,Fatin2016:PRL,Gresta2021:PRB}, which makes the field of spintronics~\cite{Zutic2004:RMP,Fabian2007:APS} also relevant in this context.

One route to probe topological superconductivity is presented by so-called topological Josephson junctions, that is, Josephson junctions made of topological superconductors. Especially topological Josephson junctions based on nanowires~\cite{Lutchyn2010:PRL,SanJose2012:PRL,Dominguez2012:PRB,Murthy2020:PRB} or on topological insulators~\cite{Fu2008:PRL,Tanaka2009:PRL,Houzet2013:PRL,Beenakker2013:PRL,Tkachov2013:PRB,Crepin2014:PRL,Tkachov2017:PRB,Tkachov2019:PRB,*Tkachov2019:JPCM,PicoCortes2017:PRB,Dominguez2017:PRB,Murani2019:PRL,Zhang2020:PRR,Keidel2020:PRR,Calzona2019:PRR,Blasi2020:PRL} have attracted a lot of interest in this context. As a hallmark of their topological nature such junctions exhibit a ground-state fermion parity that is $4\pi$-periodic in the superconducting phase difference $\varphi$ and Andreev bound states (ABS) with a protected zero-energy crossing~\cite{FuKane2009:PRB,Ioselevich2011:PRL}. The origin of this $4\pi$-periodicity and the protected crossing is that Majorana bound states localized on either side of the junction hybridize and form a peculiar protected zero-energy ABS.

Finding unambiguous experimental evidence for the protected zero-energy crossing or the $4\pi$-periodicity of the ground-state parity and related quantities proves a challenging task, however~\cite{Rokhinson2012:NP,Wiedenmann2016:NC,Laroche2019:NC,Kayyalha2019:PRL,Oostinga2013:PRX,Sochnikov2015:PRL}. For example, the Josephson current of a topological Josephson junction is $2\pi$-periodic in an equilibrium situation unless fermion parity is constrained~\cite{Beenakker2013:PRL,Tkachov2015:PRB}. Hence, typical routes to observe $4\pi$-periodic supercurrents employ non-equilibrium AC measurements driven by external fields~\cite{Rokhinson2012:NP,Wiedenmann2016:NC,Deacon2017:PRX,Laroche2019:NC}.

\begin{figure}[t]
\centering
\includegraphics*[width=8.5cm]{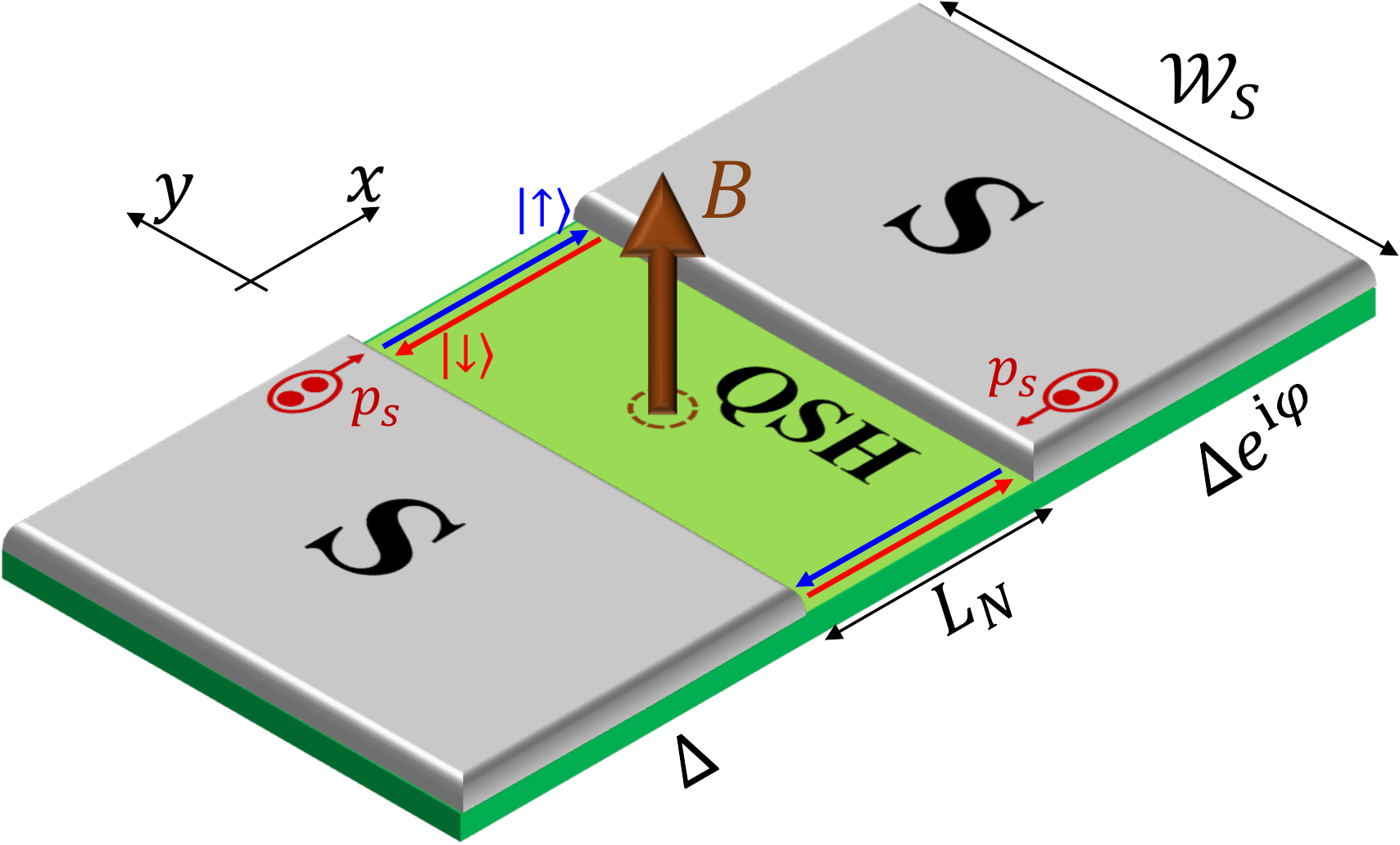}
\caption{(Color online) Scheme of a QSH-based topological Josephson junction: $s$-wave superconductors on top of the QSH insulator proximity-induce pairing of amplitude $\Delta$ into the QSH edge states $\ket{\uparrow}$ and $\ket{\downarrow}$, thus defining the superconducting (S) regions. A superconducting phase difference $\varphi$ is applied across the normal QSH weak link. In addition, a magnetic field $B$ in the QSH region induces a finite magnetic flux $B\mathcal{W}_SL_N$ and a Doppler shift as the Cooper pairs acquire a momentum $p_S$.}\label{fig:scheme}
\end{figure}

Intriguingly, thermal properties, such as the heat current driven across a Josephson junction, also depend on $\varphi$ and can provide additional insight to distinguish between trivial and topological junctions~\cite{Sothmann2016:PRB,Sothmann2017:NJP}. Furthermore, it has been shown (employing Hill thermodynamics) that topological phases exhibit unexpected universalities in their thermodynamic signatures~\cite{Kempkes2016:SR,Quelle2016:PRB}. Motivated by this and by advances in the understanding of the thermodynamics of non-topological superconductors~\cite{Giazotto2006:RMP,Rabani2008:PRB,Giazotto2012:N,Giazotto2012:APL,MartinezPerez2013:APL,Strambini2014:APL,Fornieri2016:NN,Marchegiani2016:PRAp,Paolucci2017:EPL,Fornieri2017:NN,Virtanen2017:PRB,Vischi2019:SR,Vischi2019:E}, we have recently applied these thermodynamic concepts to topological Josephson junctions to propose a topological Josephson heat engine~\cite{Scharf2020:CP}. In the same spirit, we discuss another way of utilizing coherent thermodynamics to demonstrate the peculiar nature of topological Josephson junctions: We show that measurements of the phase-dependent heat capacity can provide distinct signatures originating from the topological ABS and thus represent an alternative property which can be investigated in topological Josephson junctions. Furthermore, we explicitly propose a dispersive measurement scheme, that is, a scheme coupling the Josephson junction to a tank LC circuit via whose response one can detect the characteristic features of topological Josephson junctions experimentally, both with and without parity constraints.

As a proof-of-concept we consider a Josephson junction based on a quantum spin Hall (QSH) insulator in this paper, where the inclusion of a magnetic flux in the normal region allows for an additional, experimentally relevant, tuning parameter. The corresponding model and its ABS and continuum states are introduced in Sec.~\ref{Sec:Model}. In Secs.~\ref{Sec:DeltaNoParity} and~\ref{Sec:DeltaParity}, the thermodynamic properties of this system are discussed without and with constraints on the fermion parity, respectively. This is followed by a discussion on possible experimental realizations of measurements of the heat capacity in Sec.~\ref{Sec:Exp}. A short summary concludes the manuscript in Sec.~\ref{Sec:Conclusions}.

\section{Model, Andreev bound states and continuum states}\label{Sec:Model}
We consider a short, topological Josephson junction based on a quantum spin Hall (QSH) insulator, where the length $L_N$ of the normal (N) region is small compared to the Josephson penetration depth. In this setup, a QSH insulator is partially covered by $s$-wave superconductors, which proximity-induce pairing to the QSH edge states via tunneling and thus define the superconducting (S) regions [see Fig.~\ref{fig:scheme}]. The QSH system lies in the $xy$ plane, with the direction of the superconducting phase bias denoted as the $x$ direction. The Fermi level is situated within the bulk gap, where only edge states exist. Moreover, we assume that the two edges of the QSH system are separated by a distance $\mathcal{W}_S$ large enough so that there is no overlap between states from opposite edges. In this case, the spin projection in $z$ direction, $s=\uparrow/\downarrow\equiv\pm1$, and $\sigma=t/b\equiv\pm1$, describing the top and bottom edges, are good quantum numbers. The corresponding Bogoliubov-de Gennes (BdG) Hamiltonian reads
\begin{multline}\label{eq:BDGHamSimple}
\hat{H}_{s,\sigma}=\left(s\sigma v_F\hat{p}_x-\mu_S\right)\tau_z+s\frac{v_Fp_S}{2}+V_0h(x)\tau_z\\
+\Delta(x)\left[\tau_x\cos\Phi_\sigma(x)-\tau_y\sin\Phi_\sigma(x)\right],
\end{multline}
where $\tau_j$ (with $j=x,y,z$) denote Pauli matrices of the particle-hole degrees of freedom (Appendix~\ref{App:SimpHam}).

Here we study a short junction with a N region of length $L_N$, which we describe by a $\delta$-like N region with $h(x)=L_N\delta(x)$~\footnote{For short Josephson junctions, the Thouless energy $E_T=(\pi/2)(\hbar v_F/L_N)$ is much larger than the (proximity-induced) superconducting gap $\Delta$, that is, $\Delta\ll E_T$. For such junctions, the model of a $\delta$-like N region used in this manuscript provides an excellent description as also discussed in Appendix~\ref{App:ABSandCont}.}. In this model, the proximity-induced pairing amplitude $\Delta$~\footnote{Note that the gap $\Delta$ proximity-induced into the QSH edge states has to be clearly distinguished from the superconducting gap $\Delta_S$ of the parent superconductor. It is rather a self-energy component that for low energies becomes energy independent and is determined by the tunneling coupling strength between the QSH edge states and the parent superconductors as well as by the normal-state density of states of the parent superconductors~\cite{Tkachov2013:PRB}.} follows the constant profile $\Delta(x)=\Delta$ and the phase convention is $\Phi_{t/b}(x)=\Theta(x)\varphi_{t/b}$, where $\varphi_{t/b}$ describes the superconducting phase difference of the top and bottom edges, respectively. Furthermore, $\hat{p}_x$ denotes the momentum operator, $v_F$ the Fermi velocity of the edge states, and $V_0$ is the potential difference between the N and S regions. The presence of an (orbital) out-of-plane magnetic field $\bm{B}=B\bm{e}_z$ with $B\geq0$ provides an additional tuning parameter and has two important consequences~\cite{Tkachov2015:PRB}: $\varphi_t$ and $\varphi_b$ differ, $\varphi_{t/b}=\varphi\pm\pi B\mathcal{W}_SL_N/\Phi_0$, where $\varphi$ is the superconducting phase difference at $B=0$ and $\Phi_0$ the magnetic flux quantum~\footnote{In general, $\varphi_{t/b}$ are determined from the magnetic flux through an effective contact region $B\mathcal{W}_SL^*$, where $\mathcal{W}_S$ is the transverse width of the Josephson junction and $L^*=L_N+2\tilde{\lambda}$ is the effective length of the contact region. This effective length $L^*$ consists of the length $L_N$ of the normal region and the penetration $\tilde{\lambda}$ into each of the two proximity-induced superconducting leads. Assuming that thin superconducting films with a thickness much smaller than the parent superconductors' London penetration depth are situated on top of the QSH insulator, the penetration $\tilde{\lambda}$ is governed by the Pearl penetration depth. If the Pearl penetration depth is much larger than $\mathcal{W}_S$ one can approximate $L^*\approx L_N$~\cite{Tkachov2015:PRB}. However, we emphasize that the results presented in our work here are valid, regardless of the exact form of the magnetic flux through the contact region.}. Likewise, $\bm{B}$ induces a Doppler shift described by the Cooper pair momentum $p_S=\pi\hbar B\mathcal{W}_S/\Phi_0$. Employing a scattering approach, we determine the ABS and the continuum spectrum of Eq.~(\ref{eq:BDGHamSimple}) as detailed in Appendices~\ref{App:ABS} and~\ref{App:Continuum}.

This approach yields the ABS energies
\begin{equation}\label{eq:delta_ABS}
\epsilon^\sigma_s(\varphi_\sigma)=s\left[-\sigma\mathrm{sgn}\left(\sin\frac{\varphi_\sigma}{2}\right)\Delta\cos\frac{\varphi_\sigma}{2}+\frac{v_Fp_S}{2}\right],
\end{equation}
which describes two ABS per edge. Without loss of generality, we will from now on focus on the top edge, where the phase dependence of all quantities is governed by $\varphi_t=\varphi+\pi B\mathcal{W}_SL_N/\Phi_0$.

\begin{figure}[t]
\centering
\includegraphics*[width=8.5cm]{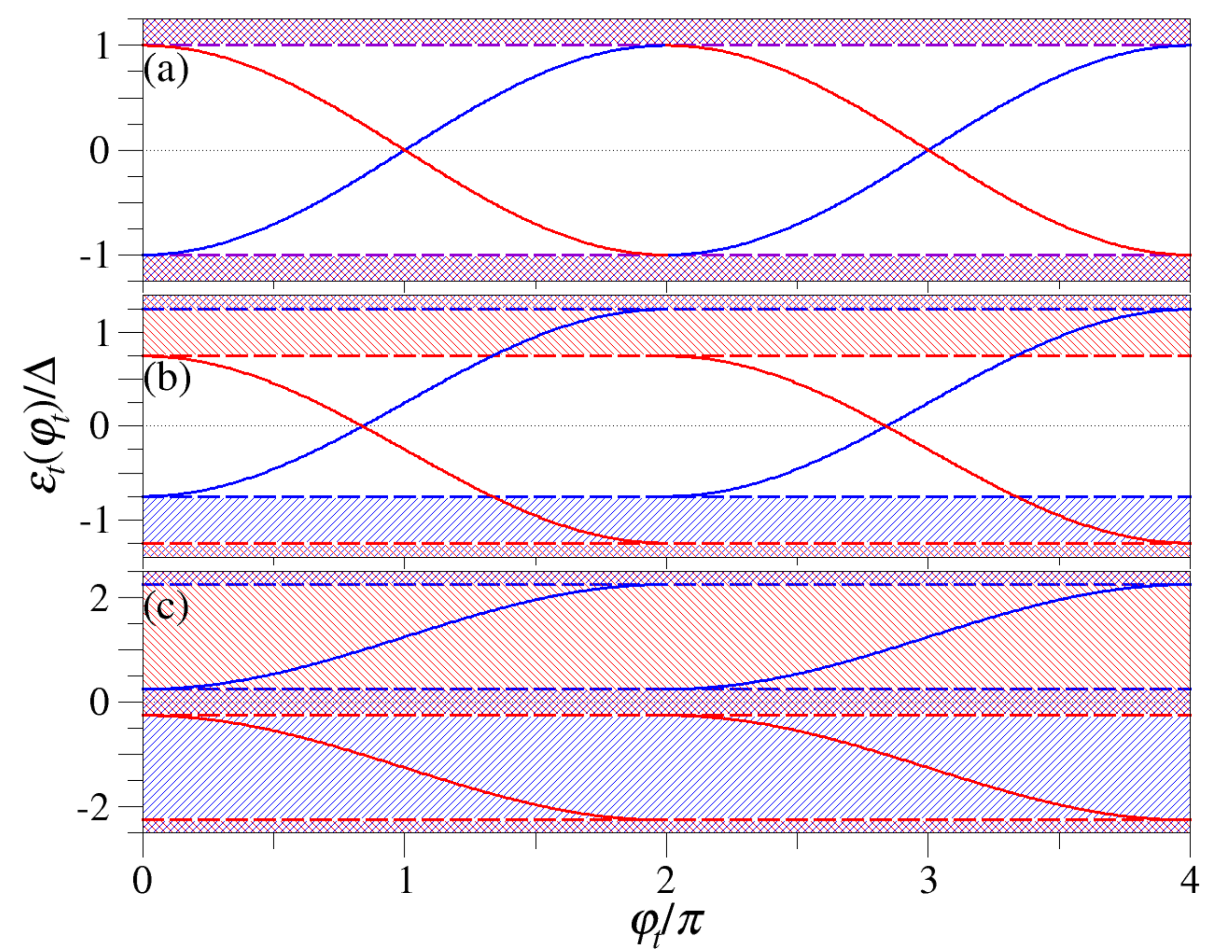}
\caption{(Color online) Andreev bound states given by Eq.~(\ref{eq:delta_ABS}) for the top edge with different $p_S$: (a) $v_Fp_S=0$, (b) $v_Fp_S=0.5\Delta$, (c) $v_Fp_S=2.5\Delta$. Here the blue and red lines denote the different spin quantum numbers $s=\uparrow/\downarrow$ and the shaded regions correspond to the continuum states.}\label{fig:ABS}
\end{figure}

Similar to Ref.~\cite{Tkachov2019:PRB}, where a closely related system has been discussed, for finite $p_S$ the dispersion~(\ref{eq:delta_ABS}) exhibits discontinuities at $\varphi_t=2\pi n$ with $n$ being an integer (Fig.~\ref{fig:ABS}). At these discontinuities, the ABS merge into the continuum states of Eq.~(\ref{eq:BDGHamSimple}) described by the continuum density of states (DOS) of the top edge,
\begin{multline}\label{eq:delta_con}
\rho_0^t(\epsilon,\varphi_t)=\frac{\Delta^2\sin\varphi_t}{2\pi}\left[\frac{\Theta\left(\epsilon_\uparrow^2-\Delta^2\right)}{\sqrt{\epsilon_\uparrow^2-\Delta^2}\left(\epsilon_\uparrow^2-\Delta^2\cos^2\frac{\varphi_t}{2}\right)}\right.\\
\left.-\frac{\Theta\left(\epsilon_\downarrow^2-\Delta^2\right)}{\sqrt{\epsilon_\downarrow^2-\Delta^2}\left(\epsilon_\downarrow^2-\Delta^2\cos^2\frac{\varphi_t}{2}\right)}\right]
\end{multline}
with $\epsilon_{\uparrow/\downarrow}=\epsilon\mp v_Fp_S/2$ (Appendix~\ref{App:Continuum}).

While these continuum states reduce the effective superconducting gap, the ABS residing in this gap still exhibit a protected crossing [Figs.~\ref{fig:ABS}(a,b)] as long as $|v_Fp_S|<2\Delta$. These crossings also correspond to changes in the ground-state fermion parity~\cite{Ioselevich2011:PRL,Beenakker2013:PRL}. If $|v_Fp_S|>2\Delta$, there are bound states, but these are all coexisting with continuum states as the superconducting gap is closed [Fig.~\ref{fig:ABS}(c)].

\section{Thermodynamics of a single edge without parity constraints}\label{Sec:DeltaNoParity}
From the ABS and continuum spectra of Eq.~(\ref{eq:BDGHamSimple}) one can calculate the free energy of the top edge, which then allows one to determine other thermodynamic quantities. First, we consider a situation where there are no parity constraints as is, for example, the case in Josephson junctions strongly coupled to external reservoirs. Then, the free energy is---up to additive $\varphi_t$-independent contributions---given by~\cite{Beenakker1992,Beenakker2013:PRL}
\begin{multline}\label{eq:FE_Delta}
F^t_0(\varphi_t,T)=-k_BT\left\{\ln\left[2\cosh\left(\frac{\epsilon^t_\uparrow(\varphi_t)}{2k_BT}\right)\right]\right.\\
\left.+\int\limits_0^\infty\d\epsilon\rho^t_c(\epsilon,\varphi_t)\ln\left[2\cosh\left(\frac{\epsilon}{2k_BT}\right)\right]\right\},
\end{multline}
where $k_B$ is the Boltzmann constant and $T$ the temperature. Equation~(\ref{eq:FE_Delta}) is composed of a first part generated by the discrete ABS and a second part due to the continuum DOS $\rho^t_c(\epsilon,\varphi_t)=\rho^t_0(\epsilon,\varphi_t)+\rho_S(\epsilon)$, consisting of the $\varphi_t$-dependent contribution from the Josephson junction given by Eq.~(\ref{eq:delta_con}) and a $\varphi_t$-independent term originating from the superconducting electrodes,
\begin{equation}\label{eq:DOSS}
\rho_S(\epsilon)=\frac{2}{\pi E_S}\frac{|\epsilon|\Theta\left(\epsilon^2-\Delta^2\right)}{\sqrt{\epsilon^2-\Delta^2}}.
\end{equation}
Here the energy scale $E_S=\hbar v_F/L_S$ is related to the total length $L_S$ of the superconducting QSH edge~\cite{Beenakker2013:PRL}.

Equation~(\ref{eq:FE_Delta}) allows us to calculate the Josephson current via~\cite{Beenakker1992}
\begin{equation}\label{eq:JoCur}
I_t(\varphi_t,T)=\frac{2e}{\hbar}\frac{\partial F_0^t(\varphi_t,T)}{\partial\varphi_t},
\end{equation}
where $e$ is the elementary charge, and the entropy via
\begin{equation}\label{eq:Entropy}
S_t(\varphi_t,T)=-\frac{\partial F_0^t(\varphi_t,T)}{\partial T}.
\end{equation}
From $S_t$ one can subsequently obtain the heat capacity
\begin{equation}\label{eq:HeatCap}
C_t(\varphi_t,T)=T\frac{\partial S_t(\varphi_t,T)}{\partial T},
\end{equation}
which is directly accessible experimentally by measuring the temperature response to a heat pulse. Intriguingly, Eqs.~(\ref{eq:JoCur}) and~(\ref{eq:Entropy}) imply that the phase dependence of $S_t$ can be determined from $I_t$ via a Maxwell relation~\cite{Virtanen2017:PRB,Vischi2019:SR,Vischi2019:E} (Appendix~\ref{App:FE}).

\begin{figure}[t]
\centering
\includegraphics*[width=8.5cm]{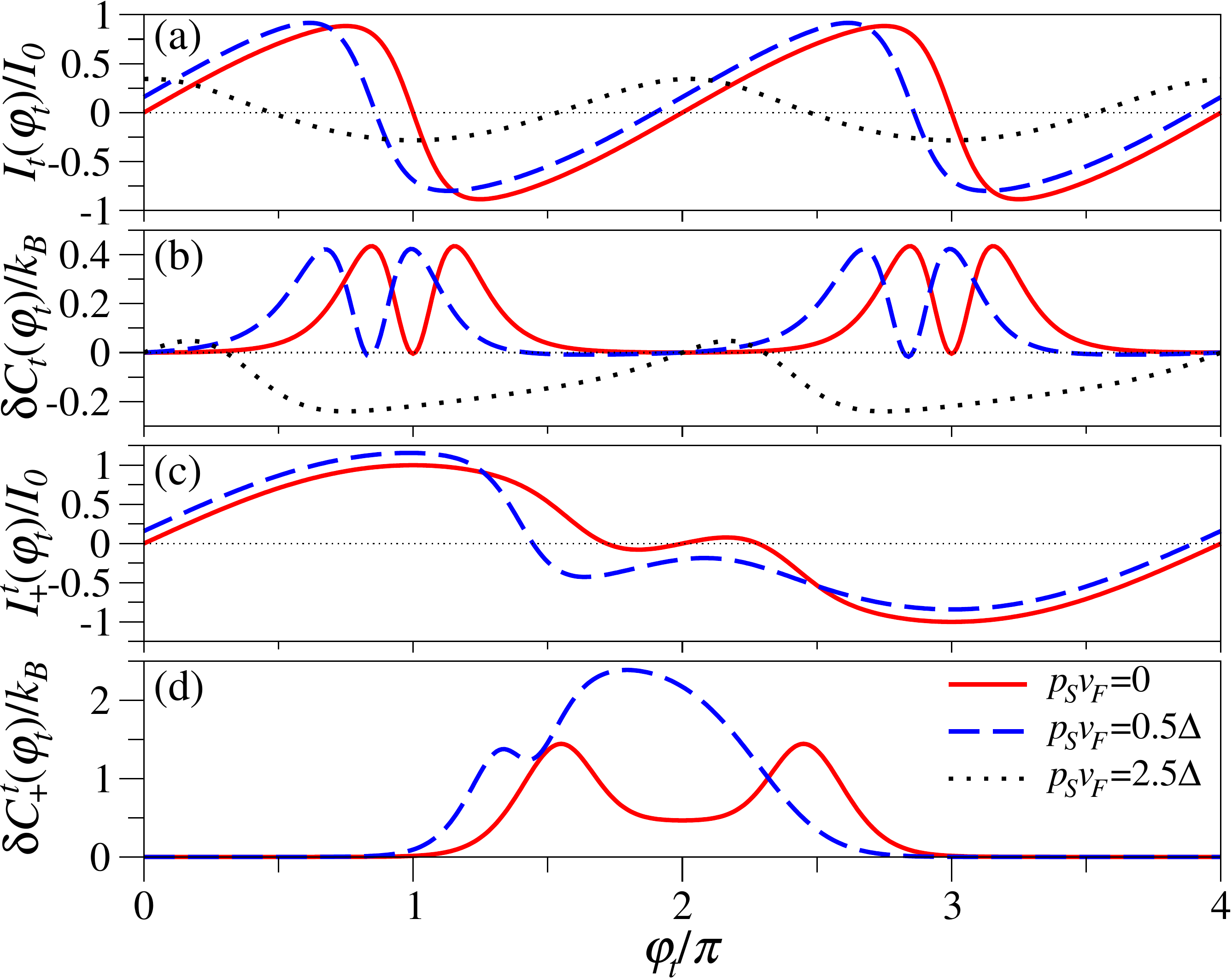}
\caption{(Color online) Phase dependence of the (a,c) Josephson current $I_t(\varphi_t)$ and of the (b,d) heat capacity $\delta C_t(\varphi_t)=C_t(\varphi_t)-C_t(0)$ for the top edge, $k_BT=0.1\Delta$ and different $p_S$ (a,b) without and (c,d) with parity constraints. Here $I_0=e\Delta/2\hbar$.}\label{fig:DeltaCHC}
\end{figure}

Without parity constraints, the only phase-dependent contributions to $F_0^t$ arise from the ABS and $\rho_0^t$. Hence, we can ignore $\rho_S$ here. Inserting Eqs.~(\ref{eq:delta_ABS}) and~(\ref{eq:delta_con}) into $F_0^t$, it is convenient to split the Josephson current, $I_t=I_{ABS}^t+I_c^t$, calculated from Eq.~(\ref{eq:JoCur}) into an ABS contribution,
\begin{multline}\label{eq:delta_JCABS}
I_{ABS}^t(\varphi_t,T)=\frac{e\Delta}{2\hbar}\sin\frac{\varphi_t}{2}\quad\quad\\
\times\tanh\left[\frac{\Delta\cos\frac{\varphi_t}{2}-\mathrm{sgn}\left(\sin\frac{\varphi_t}{2}\right)\frac{v_Fp_S}{2}}{2k_BT}\right],
\end{multline}
and a contribution from the continuum,
\begin{multline}\label{eq:delta_JCcon}
I_c^t(\varphi_t,T)=-\frac{k_BTe\Delta^2}{\pi\hbar}\int\limits_{\Delta}^{\infty}\d\epsilon\;\mathrm{ln}\left[\frac{\cosh\left(\frac{\epsilon+v_Fp_S/2}{2k_BT}\right)}{\cosh\left(\frac{\epsilon-v_Fp_S/2}{2k_BT}\right)}\right]\\
\times\frac{\epsilon^2\cos\varphi_t-\Delta^2\cos^2\frac{\varphi_t}{2}}{\sqrt{\epsilon^2-\Delta^2}\left(\epsilon^2-\Delta^2\cos^2\frac{\varphi_t}{2}\right)^2}.
\end{multline}
For $B=0$, $p_S=0$ and Eq.~(\ref{eq:delta_JCcon}) vanishes, consistent with the expectation that in short junctions the current is driven by the ABS~\cite{Beenakker1992}. Then, Eq.~(\ref{eq:delta_JCABS}) yields the well-known expression for $I_t$~\cite{Tkachov2019:PRB}. Only if $p_S\neq0$, there is a contribution from Eq.~(\ref{eq:delta_JCcon}) as shown in Fig.~\ref{fig:DeltaCHC}(a), where $I_t$ is presented as a function of $\varphi_t$ for different Doppler shifts $p_S$. For $0<|v_Fp_S|<2\Delta$, the current-phase relationship is non-sinusoidal and qualitatively similar to the case of $p_S=0$, apart from a phase shift proportional to $p_S$. As $p_S$ is increased further, $|v_Fp_S|>2\Delta$, $I_t$ becomes more sinusoidal. In this limit, in fact, the ABS contribution $I_{ABS}^t$ is unidirectional~\cite{Tkachov2019:PRB}, while the addition of the continuum contribution due to $\rho_0^t$ results in the sinusoidal behavior of $I_t$ seen in Fig.~\ref{fig:DeltaCHC}(a). For more details on the different contributions to $I_t$, we refer the reader to Appendix~\ref{App:Decomp}.

In addition to $I_t$, we can also compute $C_t$ via Eqs.~(\ref{eq:Entropy}) and~(\ref{eq:HeatCap}). We approximate $\Delta(T)\approx\Delta(T=0)$ and hence $\partial\Delta/\partial T\approx0$ when computing $C_t$, which is reasonable if $T\ll T_c$, where $T_c$ is the critical temperature of the parent superconductor. Differing from $I_t$, the exact values of $S_t$ and $C_t$ at a given $\varphi_t$ also require knowledge of the $\varphi_t$-independent contributions to the free energy~\cite{Scharf2020:CP}. To overcome this difficulty, we calculate the difference of $C_t$ with respect to its value at $\varphi_t=0$, thereby canceling the offset due to the $\varphi_t$-independent contributions. As before, it is convenient to split $C_t=C_{ABS}^t+C_c^t$ into contributions from the ABS and the continuum,
\begin{multline}\label{eq:delta_CABS}
\delta C_{ABS}^t(\varphi_t,T)=C_{ABS}^t(\varphi_t,T)-C_{ABS}^t(0,T)=\\
\frac{k_B}{\left(2k_BT\right)^2}\left\{\left[\frac{\epsilon^t_\uparrow(\varphi_t)}{\cosh\left(\frac{\epsilon^t_\uparrow(\varphi_t)}{2k_BT}\right)}\right]^2-\left[\frac{\epsilon^t_\uparrow(0)}{\cosh\left(\frac{\epsilon^t_\uparrow(0)}{2k_BT}\right)}\right]^2\right\}
\end{multline}
and
\begin{multline}\label{eq:delta_Ccon}
\delta C_c^t(\varphi_t,T)=C_c^t(\varphi_t,T)-C_c^t(0,T)=\\
\frac{k_B}{\left(2k_BT\right)^2}\int\d\epsilon\;\frac{\left[\rho_0^t(\epsilon,\varphi_t)-\rho_0^t(\epsilon,0)\right]\epsilon^2}{\cosh^2\left(\frac{\epsilon}{2k_BT}\right)},
\end{multline}
respectively.

We present the $\varphi_t$ dependence of $C_t$ for different values of $p_S$ in Fig.~\ref{fig:DeltaCHC}(b). If $p_S=0$, a double-peak feature described by Eq.~(\ref{eq:delta_CABS}) develops around the zero-energy crossing of the ABS at $\varphi_t=\pi$. Consequently, this feature remains also for $0\leq|v_Fp_S|<2\Delta$ as a hallmark of the protected ABS crossing. When $|v_Fp_S|>2\Delta$, the ABS no longer cross at zero energy as shown in Fig.~\ref{fig:ABS}(c) and thus $C_t$ no longer exhibits the double peak. The sharp double-peak feature of $C_t$ is a direct consequence of the form of Eq.~(\ref{eq:delta_CABS}), which can be shown to always possess a pronounced minimum at $\epsilon^t_\uparrow(\varphi_t)=0$, and thus provides a signature of any zero-energy crossing in the ABS. Since any short 1D topological Josephson junction exhibits an Andreev spectrum qualitatively similar to the one in Fig.~\ref{fig:ABS}(a), that is, a protected zero-energy crossing without additional ABS, the phase dependence of the heat capacity for these junctions is also qualitatively similar to the one shown in Fig.~\ref{fig:DeltaCHC}(b)~\footnote{For topological Josephson junction with multiple ABS, such as, for example, a 2D topological Josephson junction based on the surface states of a 3D topological insulator~\cite{Fu2008:PRL,Tanaka2009:PRL}, on the other hand, all ABS contribute to the heat capacity. Consequently, the contributions from the unprotected ABS superimpose the contribution of the protected zero-energy mode in the heat capacity.}.

The double-peak feature of $C_t$ thus allows one to distinguish between topological and trivial junctions: For a phase difference of $\pi$, the ABS of a trivial junction exhibit a gap around zero energy, the size of which is controlled by the potential difference $V_0$ between the S and N regions (which in turn controls the transparency of the junction). At low $T$, this gap, in turn, gives rise to a single peak of the heat capacity (Appendix~\ref{App:TopVsTriv}). On the other hand, $V_0$ cannot remove the zero-energy crossing of the ABS in a topological junction, as can be seen from Eq.~(\ref{eq:delta_ABS}), which like Eq.~(\ref{eq:delta_con}) does not depend on $V_0$~\footnote{Since the protected zero-energy crossing is the origin of the double peak, this characteristic feature of the heat capacity appears even in other topological Josephson junctions where spin is not a good quantum number. The only necessary requirement for the double-peak feature to occur for any $V_0$ is a topologically protected zero-energy crossing.}. Hence, tuning $V_0$ and monitoring the phase dependence of the heat capacity always yields a double peak in topological junctions, whereas only a single peak arises in trivial junctions as $V_0$ increases.

While modulating $V_0$ thus provides a route to test the protected crossing via $C_t$, the above quantities all exhibit a $2\pi$-periodicity with $\varphi_t$. In order to observe a $4\pi$-periodicity in the thermodynamic quantities, external constraints on the fermion parity have to be imposed.

\section{Thermodynamics of a single edge with parity constraints}\label{Sec:DeltaParity}
As long as $|v_Fp_S|<2\Delta$, there is a crossing and one can define the ground-state parity of the top edge by
\begin{equation}\label{eq:DefParity}
P(\varphi_t)=\mathrm{sgn}\left[\cos\frac{\varphi_t}{2}-\frac{v_Fp_S}{2\Delta}\mathrm{sgn}\left(\sin\frac{\varphi_t}{2}\right)\right].
\end{equation}
In defining Eq.~(\ref{eq:DefParity}), we have chosen the convention that $P(\varphi_t=0)=+1$. If the fermion parity is kept constant, the free energy acquires an additional contribution and reads~\cite{Ioselevich2011:PRL,Beenakker2013:PRL}
\begin{widetext}
\begin{equation}\label{eq:deltaFParity}
F^t_p(\varphi_t,T)=F^t_0(\varphi_t,T)-k_BT\ln\left\{\frac{1}{2}\left[1+pP(\varphi_t)\tanh\left|\frac{\epsilon^t_\uparrow(\varphi_t)}{2k_BT}\right|\exp\left[J_S(T)+\int\limits_0^\infty\d\epsilon\rho^t_0(\epsilon,\varphi_t)\ln\left(\tanh\left(\frac{\epsilon}{2k_BT}\right)\right)\right]\right]\right\}
\end{equation}
\end{widetext}
with $p$ corresponding to the lower ($p=+1$) and upper ($p=-1$) branches at $\varphi_t=0$, respectively. Again, we omit additive $\varphi_t$-independent contributions to $F^t_p$ in Eq.~(\ref{eq:deltaFParity}). The contribution
\begin{equation}\label{eq:Js}
J_S(T)=-\frac{2}{\pi k_BTE_S}\int\limits_\Delta^\infty\d\epsilon\;\frac{\sqrt{\epsilon^2-\Delta^2}}{\sinh\left(\epsilon/k_BT\right)}
\end{equation}
originates from the superconducting electrodes and can be approximated by $J_S(T)\approx-(4\Delta/\pi E_S)K_1(\Delta/k_BT)$ for $k_BT\ll\Delta$, where $K_1$ is the modified Bessel function of the second kind.

We now have to evaluate Eq.~(\ref{eq:deltaFParity}) using the ABS and $\rho^t_c$. The resulting expression for $F_p^t$ is in general quite cumbersome and we do not provide it here explicitly. For the case of $p_S=0$, however, the problem is simplified significantly because $\rho^t_0\to0$ and
\begin{multline}\label{eq:deltaFParity0}
F_p^t(\varphi_t,T)=\quad\quad\\
-k_BT\ln\left[\cosh\left(\frac{\Delta\cos\frac{\varphi_t}{2}}{2k_BT}\right)+p\e^{J_S}\sinh\left(\frac{\Delta\cos\frac{\varphi_t}{2}}{2k_BT}\right)\right].
\end{multline}
The different thermodynamic quantities under parity constraints can again be obtained from Eqs.~(\ref{eq:JoCur})-(\ref{eq:HeatCap}) if we replace $F_0^t$ by $F_p^t$. For the corresponding Josephson current, we obtain
\begin{multline}\label{eq:deltaJCParity0}
I_p^t(\varphi_t,T)=\\
\frac{e\Delta}{2\hbar}\sin\frac{\varphi_t}{2}\frac{\sinh\left(\frac{\Delta\cos\frac{\varphi_t}{2}}{2k_BT}\right)+p\e^{J_S}\cosh\left(\frac{\Delta\cos\frac{\varphi_t}{2}}{2k_BT}\right)}{\cosh\left(\frac{\Delta\cos\frac{\varphi_t}{2}}{2k_BT}\right)+p\e^{J_S}\sinh\left(\frac{\Delta\cos\frac{\varphi_t}{2}}{2k_BT}\right)}.
\end{multline}
If $J_S\to0$, which is, for example, the case for $T\to0$, we recover $I_p^t=p(e\Delta/2\hbar)\sin(\varphi_t/2)$~\cite{FuKane2009:PRB,Beenakker2013:PRL}. Due to the $T$ dependence of $J_S$, the expressions for $S_p^t$ and $C_p^t$ turn out to be very cumbersome even for $p_S=0$.

As $F_p^t$ exhibits a $4\pi$-periodicity with $\varphi_t$, $I_p^t$ and $C_p^t$ inherit this periodicity as illustrated by Figs.~\ref{fig:DeltaCHC}(c,d), where we have chosen the parity branch $p=+1$. This $4\pi$-periodicity is in contrast to the behavior of the thermodynamic quantities without parity constraints (Sec.~\ref{Sec:DeltaNoParity}). Therefore, measurements of, for example, the heat capacity offer additional possibilities to confirm the $4\pi$-periodicity in $\varphi_t$. As shown in Figs.~\ref{fig:DeltaCHC}(c,d) this $4\pi$-periodicity can also be observed for finite $|v_Fp_S|<2\Delta$. For $|v_Fp_S|>2\Delta$, on the other hand, we can no longer impose any parity constraints on the ABS.

We conclude our discussion of the thermodynamic properties of a single edge by mentioning that we have provided expressions for the top edge in Secs.~\ref{Sec:DeltaNoParity} and~\ref{Sec:DeltaParity}. The corresponding expressions for the bottom edge are given by Eqs.~(\ref{eq:delta_con})-(\ref{eq:deltaJCParity0}) with the substitution $B\to-B$, that is, $\varphi_t\to\varphi_b$, $p_S\to-p_S$, $\epsilon_\downarrow\to\epsilon_\uparrow$ and $\epsilon_\uparrow\to\epsilon_\downarrow$.

\begin{figure}[t]
\centering
\includegraphics*[width=\linewidth]{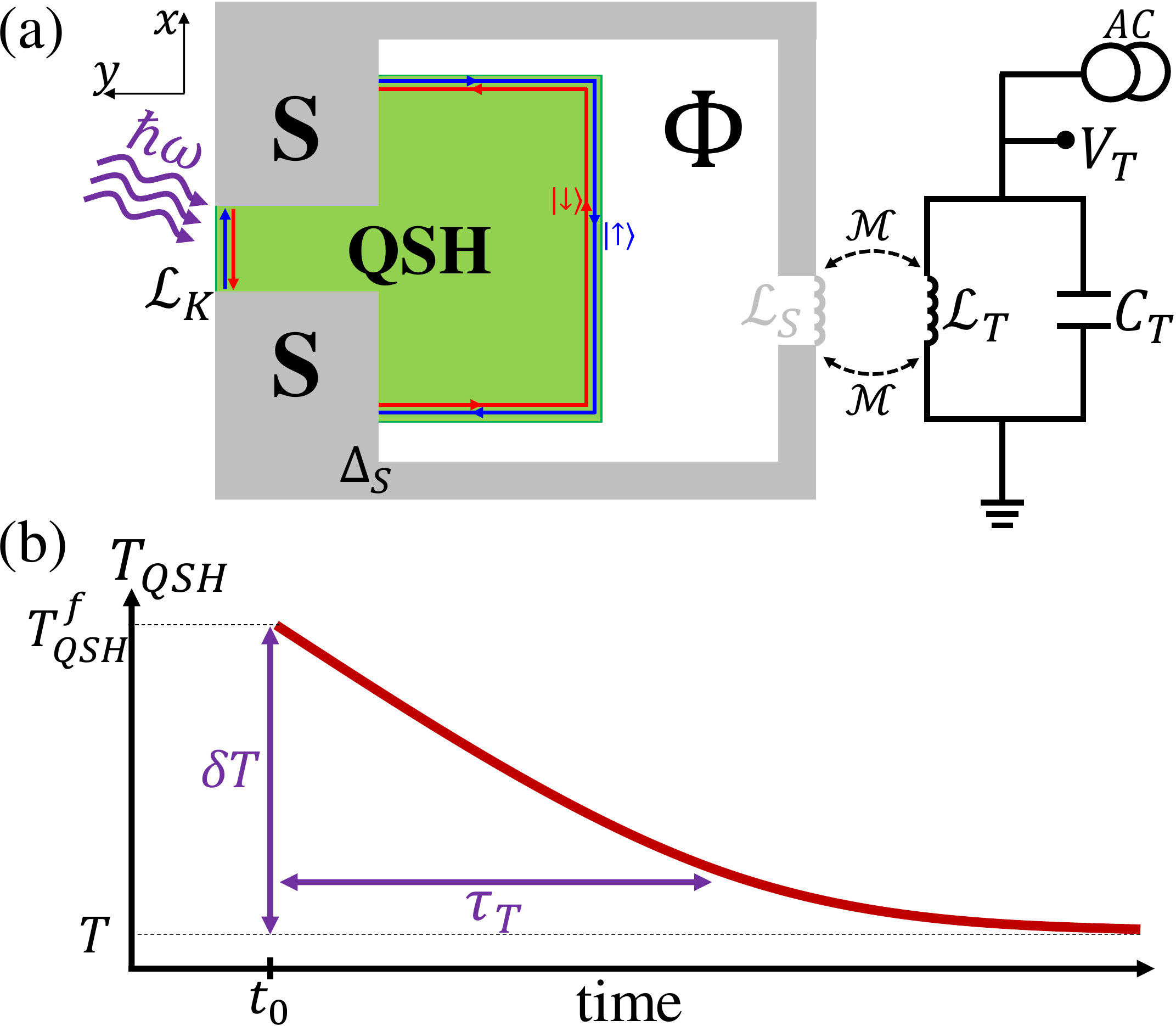}
\caption{(Color online) (a) Setup to experimentally measure the heat capacity: A circuit consisting of a topological Josephson junction (with kinetic inductance $\mathcal{L}_K$) inside a superconducting ring (with superconducting gap $\Delta_S$ and inductance $\mathcal{L}_S$) threaded by a magnetic flux $\Phi$ is coupled via the mutual inductance $\mathcal{M}$ to a tank circuit composed of an inductor $\mathcal{L}_T$ and a capacitor $C_T$. The tank circuit is connected to an AC current source and shunted to a voltage probe $V_T$. By measuring the dispersive response of the resonant circuit it is possible to monitor the temperature $T_{QSH}$. The topological Josephson junction is subject to a light pulse with photon energies $\hbar\omega<\Delta_S$. (b) Time evolution of $T_{QSH}$ after the light pulse at time $t_0$: The pulse generates an increase $\delta T$ in temperature compared to the bath temperature $T$. After the pulse, the temperature decays back to the bath temperature with a characteristic time scale $\tau_T$.}\label{fig:Exp}
\end{figure}

\section{Experimental realizations}\label{Sec:Exp}
Finally, we discuss a potential experimental realization to measure the heat capacity $C_t(\varphi_t,T)$ of one (top) QSH edge of the topological junction. Measuring small heat capacities requires a detection scheme similar to low-energy calorimetry~\cite{Denlinger1994:RSI,Fominaya1997:RSI,Marnieros1999:PB,Lindell2000:PB,Bourgeois2005:PRL,Guarcello2019:PRAp}. As the measurement should be conducted quite fast (see below), we also take inspiration from fast thermometry techniques~\cite{Zgirski2018:PRAp,Wang2017:APL}. Our basic idea here is to subject the topological Josephson junction to radiative heating~\cite{Pendry1999:JPCM,Qiu1993:JHT,Baffou2013:LPR}: An incident electromagnetic radiation with photon energies $\hbar\omega$ is absorbed by the junction. When $\hbar\omega<\Delta_S$, where $\Delta_S$ is the gap of the parent superconductor S, the radiation is mainly absorbed at the proximitized single-edge Josephson junction as shown in Fig.~\ref{fig:Exp}. For sufficiently short Josephson junctions, such as considered here, the time of flight $\tau_f=L/v_F$ could be very short, of the order of $10^{-11}$s, if the total junction length $L$ is a few micrometers. Hence, for time scales $t\gtrsim\tau_f$, one can define a temperature $T_{QSH}$ for the junction. 

We assume a single-edge topological Josephson junction irradiated with a low-intensity flux of photons such that the time between single-photon absorption events is longer than the dead time, that is, the time required for the junction to be restored to its initial state. During the absorption of the single-photon energy $\hbar\omega$ one observes an increase $\delta T$ in the temperature such that the final temperature becomes $T_{QSH}^f=T+\delta T$, with the bath temperature $T$ assumed to be the equilibrium temperature before absorption. The increase of the temperature can be estimated as~\cite{Guarcello2019:PRAp}
\begin{equation}
\label{eq:LightPulse}
\int_{T}^{T+\delta T}\!\!\!\! dT_{QSH}\ C_t(\varphi_t,T_{QSH})=\hbar\omega\,.
\end{equation}
For sufficiently small photon energies $\hbar\omega$, it can be approximated as $\delta T\approx\hbar\omega/C_t(\varphi_t,T)$, with the increase of temperature inversely proportional to the thermal capacity. In Fig.~\ref{fig:Exp}(b), we represent the behavior of $T_{QSH}$ after absorption of the photon energy $\hbar\omega$ at $t=t_0$. The junction relaxes back to the bath temperature $T$ with a characteristic time scale $\tau_T$ (which here plays the role of the dead time) [see Fig.~\ref{fig:Exp}(b)]. This time scale is mainly determined  by the strengths of the different mechanisms which restore thermal equilibrium such as electron-phonon coupling, radiative losses and heat leakage through the lateral S banks. In the relaxation dynamics, the photon energy $\hbar\omega$, which is initially converted fast to thermal energy, is progressively lost to the environment due to thermal losses. Similarly to what happens during the discharging of a capacitor due to a load in an electrical analogy, the thermal response time can also be connected to the thermal capacity~\cite{Guarcello2019:PRAp} via $\tau_T\approx C_t(\varphi_t,T)/G_{th}$, where $G_{th}$ is the thermal conductance describing all the different thermal losses in the junction. Although this appears to show another way to measure the heat capacitance, it is hard to think of measuring the thermal capacity from $\tau_T$ since this requires detailed knowledge of the thermal losses $G_{th}$ of the topological Josephson junction, something that---to the best of our knowledge---still needs to be investigated theoretically and experimentally~\footnote{Incidentally, the proposed setup could potentially also be used to measure those unknown quantities}. The relaxation mechanism could become quite slow however. Indeed, in a Josephson junction the thermal losses are reduced due to the thermal opacity of the superconductors, that is, $G_{th}$ is expected to be small. This time scale could become even longer than the time scale for quasiparticle poisoning (see below). Hence, we propose to measure the thermal capacitance only using the first method we have discussed.

In order to make the measurement of $\delta T$, one has to perform a fast and precise measurement of the temperature $T_{QSH}$ on time scales $t<\tau_T$. Taking inspiration from fast thermometry techniques~\cite{Zgirski2018:PRAp,Wang2017:APL}, we propose the setup shown in Fig.~\ref{fig:Exp}(a). The topological Josephson junction is inserted in a superconducting ring where the superconducting phase difference is mainly driven by the flux $\Phi$~\footnote{Note that the magnetic field which generates the flux $\Phi$ also affects the Doppler shift. Since the effective area of the superconducting ring is expected to be much larger than the small surface of the topological junction, we expect that $\Phi$ can be manipulated through small variations of the magnetic field without significantly affecting the Doppler shift.}. This resembles the configuration of an rf-SQUID for the topological junction. The superconducting ring with inductance $\mathcal{L}_S$ is coupled with a mutual inductance $\mathcal{M}$ to a tank LC circuit. The resonant frequency of the LC circuit, $f=(2\pi \tilde{\mathcal{L}}_TC_T)^{-1}$, is affected by the coupling to the superconducting ring. In particular, the effective tank circuit inductance $\tilde{\mathcal{L}}_T$ becomes~\cite{Guarcello2018:PRAp}
\begin{equation}\label{eq:TankInduct}
\tilde{\mathcal{L}}_T=\mathcal{L}_T\left[1-\frac{\mathcal{M}^2}{\mathcal{L}_T}\frac{1}{\mathcal{L}_K(\Phi, T_{QSH})+\mathcal{L}_S}\right],
\end{equation}
renormalizing the intrinsic tank inductance $\mathcal{L}_T$ because of the superconducting ring, where the inductance is dominated by the kinetic inductance $\mathcal{L}_K(\Phi,T)$ of the topological Josephson junction [see Fig.~\ref{fig:Exp}(a)] since $\mathcal{L}_K\gg\mathcal{L}_S$. The kinetic inductance of the Josephson junction is 
\begin{equation}\label{eq:KinInduct}
\mathcal{L}_K=\frac{\Phi_0}{2\pi}\left(\frac{\partial I_t(\varphi_t,T)}{\partial\Phi}\right)^{-1},
\end{equation}
where $I_t(\varphi_t,T)$ is the Josephson current of the junction computed from Eq.~(\ref{eq:JoCur}) as in the analysis in Secs.~\ref{Sec:DeltaNoParity} and~\ref{Sec:DeltaParity}. In this setup, the temperature measurement is done in a dispersive way, that is, the LC tank circuit resonance can be investigated with a small rf-signal which probes the resonance. Indeed, any change of the junction temperature $T_{QSH}$ will affect the Josephson current $I_t(\varphi_t,T_{QSH})$, the kinetic inductance $\mathcal{L}_K$ and finally the resonance frequency $f_T$ of the tank circuit. Following a calibration stage where the characteristics of the resonance spectrum are investigated in terms of the temperature of the junction, one can define the optimal strategy and setup for fast measurements of the junction temperature~\footnote{Note that this calibration stage returns also direct information on the Josephson current in itself with the potential to cross-validate the analysis from different perspectives.}. Now, one needs to relate the measurable temperature $T_{QSH}$ and its radiation-induced change $\delta T$ to the heat capacity of the QSH edges.

It is useful to provide a rough estimate of the expected temperature increase in the proposed setup. We first assume a realistic value for the proximitized gap of the topological junction $\Delta\approx 40$ $\mu$eV, corresponding to 500 mK~\cite{Bocquillon2018,Ren2019:N}, and a bath temperature of $T= 0.1\Delta$, corresponding to around 50 mK, which represents exactly the case considered in Secs.~\ref{Sec:DeltaNoParity} and~\ref{Sec:DeltaParity}. For a (proximitized or intrinsic) superconductor of gap $\Delta$ the thermal capacitance at low temperature can be written as $C_S(T)\approx k_B N_0\sqrt{2\pi} (\Delta/k_BT)^{3/2}e^{-\Delta/k_BT}\Delta$ with $N_0$ the superconductor density of states~\cite{Abrikosov1963,Vischi2019:SR}. This analytical form shows that the thermal capacity is exponentially suppressed with decreasing temperature as $\propto e^{-\Delta/k_BT}$. Thus, even including the contribution from the parent superconductors taken as Al films with a superconducting gap of 200 $\mu$eV, the phase-independent term of the thermal capacity hardly surpasses a few $k_B$ at such low temperatures. Looking at Fig.~\ref{fig:DeltaCHC}, one can discern that the phase-dependent term of the thermal capacitance, $\delta C_t(\varphi_t,T)=C_t(\varphi_t,T)-C_t(0,T)$, contributes with a comparable amount $\sim k_B$. Assuming photon absorption in the range of $\omega\sim 2$ GHz, one finds a temperature increase of $\delta T\sim10-100$ mK, having assumed a thermal capacitance of just $C_t(\varphi_t,T)\sim 1-10 k_B$. These temperature differences are appreciable with respect to the fixed thermal bath temperature $T$. It is crucial in this setup to observe that the flux $\Phi$ fixes the superconducting phase difference between the Josephson junction, fixing $\varphi_t$, and that the temperature calibration has to be done for every value of $\Phi$. This technique can thus be utilized to study the full phase dependence of $C_t(\varphi_t,T)$ with its characteristic features discussed in Secs.~\ref{Sec:DeltaNoParity} and~\ref{Sec:DeltaParity}, for standard and constrained thermodynamics, respectively.

Finally to measure the $4\pi$-periodic heat capacity of a system with fermion parity constraints, the measurement of $C_t(\varphi_t,T)$ needs to be done faster than the quasiparticle poisoning time. With typical quasiparticle poisoning times of the order of 1 $\mu$s~\cite{Virtanen2013:PRB,Frombach2020:PRB} and a resonant frequency of the tank circuit of a few GHz, we expect a temperature detection on the scale of ns or even less. Such time scales can in principle be reached as demonstrated in qubit and SQUID technology~\cite{Mueck2010:SST}. Hence, even measuring a $4\pi$-periodic heat capacity appears feasible~\cite{Scharf2020:CP}. However, even without conserving the ground-state fermion parity, the phase dependence of the heat capacity exhibits pronounced signatures originating from the topological nature of the Josephson junction studied here (see Sec.~\ref{Sec:DeltaNoParity}).

\section{Conclusions}\label{Sec:Conclusions}
In this work, we have analyzed key properties of the thermodynamics in topological Josephson junctions based on quantum spin Hall insulators. The $4\pi$-periodicity of the free energy in the phase bias $\varphi$ is replicated in other thermodynamic observables when fermion parity is conserved. As a consequence, measuring the $\varphi$ dependence of the heat capacity offers an intriguing alternative to confirm the topological nature of the Andreev bound states. If, on the other hand, there is no fermion parity conservation, the thermodynamic observables will exhibit only a $2\pi$-periodicity in $\varphi$. Even in this case, however, the protected zero-energy crossing of the Andreev spectrum manifests itself in a sharp double-peak feature in the $\varphi$ dependence of the heat capacity. Since this double-peak feature is robust against changes in the strength of the tunneling barrier only in a topological junction, one can distinguish between topological and non-topological Josephson junctions by modulating the tunneling-barrier strength, for example, via a gate on top of the normal region.

The inclusion of an out-of-plane magnetic flux inside the normal region induces a Doppler shift and allows for an additional tuning parameter to control the protected zero-energy crossing in the Andreev spectrum: While this Doppler shift reduces the effective gap of the Andreev spectrum, the crossing remains as long as this gap is not completely closed. The gap closing and disappearance of the zero-energy crossing in the Andreev spectrum with increasing magnetic flux is also reflected in the heat capacity. Although our results have been obtained for a quantum-spin-Hall-based junction, they are also applicable to other topological Josephson junctions, such as nanowire-based junctions.

Finally, we have also provided a dispersive setup to measure the small thermal capacity at fixed phase bias which is sufficiently fast to detect the $4\pi$-periodicity in the absence of quasiparticle poisoning. Comparison with the $2\pi$-periodic equilibrium heat capacity also provides a method to indirectly investigate thermal losses and realize novel calorimetric applications. 

\acknowledgments
B.S. and E.M.H. acknowledge funding by the Deutsche Forschungsgemeinschaft (DFG, German Research Foundation) through SFB 1170, Project-ID 258499086, through Grant No. HA 5893/4-1 within SPP 1666 and through the W\"urzburg-Dresden Cluster of Excellence on Complexity and Topology in Quantum Matter -- \textit{ct.qmat} (EXC 2147, Project-ID 390858490) as well as by the ENB Graduate School on Topological Insulators. E.S., A.B. and F.G acknowledge partial financial support from the EU’s Horizon 2020 research and innovation program under Grant Agreement No. 800923 (SUPERTED), the CNR-CONICET cooperation program "Energy conversion in quantum nanoscale hybrid devices", the SNS-WIS jointlab QUANTRA funded by the Italian Ministry of Foreign Affairs and International Cooperation and the Royal Society through the International Exchanges between the UK and Italy (Grant No.IEC R2192166).

\appendix

\begin{widetext}

\section{Simplified Hamiltonian}\label{App:SimpHam}
In this section, we provide additional details on the derivation of the simplified BdG Hamiltonian~(\ref{eq:BDGHamSimple}), which serves as the starting point of our calculations. With the basis order $\left(\hat{\psi}_{t,\uparrow},\hat{\psi}_{t,\downarrow},\hat{\psi}_{b,\uparrow},\hat{\psi}_{b,\downarrow},\hat{\psi}^\dagger_{t,\downarrow},-\hat{\psi}^\dagger_{t,\uparrow},\hat{\psi}^\dagger_{b,\downarrow},-\hat{\psi}^\dagger_{b,\uparrow}\right)$, the BdG Hamiltonian describing the Josephson junction introduced in Sec.~\ref{Sec:Model} is given by
\begin{multline}\label{eq:BDGHam}
\hat{H}_\mathrm{BdG}=\left(v_F\hat{p}_xs_z\sigma_z-\mu_S\right)\tau_z+\frac{v_Fp_S}{2}s_z+V_0h(x)\tau_z\\
+\Delta(x)\frac{\sigma_0+\sigma_z}{2}\left[\tau_x\cos\Phi_t(x)-\tau_y\sin\Phi_t(x)\right]+\Delta(x)\frac{\sigma_0-\sigma_z}{2}\left[\tau_x\cos\Phi_b(x)-\tau_y\sin\Phi_b(x)\right].
\end{multline}
Here $s_j$, $\sigma_j$ and $\tau_j$ (with $j=x,y,z$) denote Pauli matrices describing spin, top/bottom edge and particle-hole degrees of freedom, respectively. Note that unit matrices are not written explicitly in Eq.~(\ref{eq:BDGHam}). The N region is described by the profile $h(x)$ and the potential $V_0$, which can also be viewed as describing the difference between the chemical potentials in the S and N regions, $\mu_S$ and $\mu_N=\mu_S-V_0$. Here we use a $\delta$-barrier model with $h(x)=L_N\delta(x)$ and $\Delta(x)=\Delta$. A more general approach to the junction would be to use a finite N region with $h(x)=\Theta(L_N-x)\Theta(x)$ and $\Delta(x)=\Delta\left[\Theta(x-L_N)+\Theta(-x)\right]$. Such a model is capable of describing both a short as well as a long junction. As explained in Appendix~\ref{App:ABSandCont} below, the $\delta$-barrier model is still suitable to capture the essential physics of a short junction based on QSH edge states.

Since $\left[\hat{H}_\mathrm{BdG},s_z\right]=\left[\hat{H}_\mathrm{BdG},\sigma_z\right]=0$, the wave functions can be described by the good quantum numbers $s=\uparrow/\downarrow\equiv\pm1$ for the spin projection in $z$ direction and $\sigma=t/b\equiv\pm1$ for the top and bottom edges. Hence, an eigenstate $\Psi_{s,\sigma}(x)$ of the BdG Hamiltonian~(\ref{eq:BDGHam}) can be written as $\Psi_{s,\sigma}(x)=\psi_{s,\sigma}(x)\otimes\eta_\sigma\otimes\chi_s$, which satisfies
\begin{equation}\label{eq:BDG}
\hat{H}_{s,\sigma}\psi_{s,\sigma}(x)=E\psi_{s,\sigma}(x),
\end{equation}
where $\hat{H}_{s,\sigma}$ is given by Eq.~(\ref{eq:BDGHamSimple}) in the main text. The spinors $\chi_s$ and $\eta_\sigma$ are the eigenvectors of the Pauli matrices $s_z$ and $\sigma_z$, respectively. They satisfy $s_z\chi_s=s\chi_s$ and $\sigma_z\eta_\sigma=\sigma\eta_\sigma$.

\section{Computing the Andreev and continuum spectra}\label{App:ABSandCont}

In order to compute the ABS as well as the $\varphi$-dependent part of the continuum DOS $\rho_c(\epsilon,\varphi)$, we employ a wave-function matching approach to solve Eq.~(\ref{eq:BDG}). For the following derivations, it is convenient to introduce the spin-dependent energy variable $\epsilon_s=\epsilon-sv_Fp_S/2$.

\subsection{Andreev bound states}\label{App:ABS}
For ABS, that is, states with $|\epsilon_s|<\Delta$ and thus states localized around the N region, we can make the piecewise ansatz
\begin{equation}\label{eq:ABSansatz}
\psi_{s,\sigma}(x)=\frac{1}{\sqrt{L_{tot}}}\left\{\begin{array}{l}
a_{s,\sigma}\left(\begin{array}{l}1\\C^{(s)}_{s\sigma}\end{array}\right)\e^{\i s\sigma k_Fx}\e^{\kappa^{(s)}_\epsilon x},\; x<0,\\
\quad\\
\quad\\
e_{s,\sigma}\left(\begin{array}{l}1\\0\end{array}\right)\e^{\i s\sigma(k_N+k^{(s)}_\epsilon)x}+h_{s,\sigma}\left(\begin{array}{l}0\\1\end{array}\right)\e^{\i s\sigma(k_N-k^{(s)}_\epsilon)x},\; 0<x<L_N,
\quad\\
\quad\\
b_{s,\sigma}\left(\begin{array}{l}1\\C^{(s)}_{-s\sigma}\e^{-\i\varphi_\sigma}\end{array}\right)\e^{\i s\sigma k_Fx}\e^{-\kappa^{(s)}_\epsilon x},\; x>L_N
\end{array}\right.
\end{equation}
for a junction with a finite N region. The ansatz for a $\delta$-junction is similar to Eq.~(\ref{eq:ABSansatz}), with the states $\psi_{s,\sigma}(x<0)$ given by the first line of Eq.~(\ref{eq:ABSansatz}) and $\psi_{s,\sigma}(x>0)$ given by the third line of Eq.~(\ref{eq:ABSansatz}). In Eq.~(\ref{eq:ABSansatz}), $C^{(s)}_\pm=(\epsilon_s\pm\i\sqrt{\Delta^2-\epsilon_s^2})/\Delta$, $k_F=\mu_S/(\hbar v_F)$, $\kappa^{(s)}_\epsilon=\sqrt{\Delta^2-\epsilon_s^2}/(\hbar v_F)$, $k_N=\mu_N/(\hbar v_F)$, $k^{(s)}_\epsilon=\epsilon_s/(\hbar v_F)$, and $L_{tot}=L_N+L_S$ is the unit length of the entire S/N/S edge. The ansatz~(\ref{eq:ABSansatz}) for $\psi_{s,\sigma}(x)$ has been chosen in such a way that Eq.~(\ref{eq:BDG}) is satisfied in each S and N region separately and that $\lim\limits_{x\to\pm\infty}\psi_{s,\sigma}(x)=0$. The coefficients $a_{s,\sigma}$, $b_{s,\sigma}$, $e_{s,\sigma}$, and $h_{s,\sigma}$ have to be determined from the boundary conditions at the S/N interfaces.

For the $\delta$-barrier, the boundary condition can be obtained by integrating Eq.~(\ref{eq:BDG}) from $x=-\eta$ to $x=\eta$ with $\eta\to0^+$. The corresponding procedure~\cite{MatosAbiague2003:PRA,Sothmann2016:PRB,Scharf2016:PRL,Maistrenko2021:PRB} yields
\begin{equation}\label{eq:BCdelta}
\psi_{s,\sigma}(0^+)=\e^{-\i s\sigma Z_0}\psi_{s,\sigma}(0^-),
\end{equation}
where $Z_0=V_0L_N/(\hbar v_F)$. For a finite barrier, on the other hand, the boundary conditions require $\psi_{s,\sigma}(x)$ to be continuous at the S/N interfaces,
\begin{equation}\label{eq:BCfinite}
\psi_{s,\sigma}(0^+)=\psi_{s,\sigma}(0^-),\quad\psi_{s,\sigma}(L_N^+)=\psi_{s,\sigma}(L_N^-).
\end{equation}

First, we consider the $\delta$-barrier model, for which we invoke the boundary condition~(\ref{eq:BCdelta}) and require a nontrivial solution for the coefficients $a_{s,\sigma}$ and $b_{s,\sigma}$. This procedure yields a transcendental equation for the energy variable $\epsilon_s$ in the form of
\begin{equation}\label{eq:transABSdelta}
\arccos\left(\frac{\epsilon_s}{\Delta}\right)=-s\sigma\frac{\varphi_\sigma}{2}+\pi n,
\end{equation}
where $n$ is an integer that has to be chosen in such a way that $0\leq\arccos\left(\epsilon_s/\Delta\right)\leq\pi$. Solving Eq.~(\ref{eq:transABSdelta}) and using $\epsilon_s=\epsilon-sv_Fp_S/2$, we obtain the ABS given by Eq.~(\ref{eq:delta_ABS}) in the main text.

For a finite N region, we use the boundary conditions~(\ref{eq:BCfinite}) and require a nontrivial solution for the coefficients $a_{s,\sigma}$, $b_{s,\sigma}$, $e_{s,\sigma}$, and $h_{s,\sigma}$. Now, we obtain the transcendental equation
\begin{equation}\label{eq:transABSfinite}
\arccos\left(\frac{\epsilon_s}{\Delta}\right)-\frac{\epsilon_sL_N}{\hbar v_F}=-s\sigma\frac{\varphi_\sigma}{2}+\pi n,
\end{equation}
where $n$ is again an integer chosen in such a way that $0\leq\arccos\left(\epsilon_s/\Delta\right)\leq\pi$. The second term on the left-hand side of Eq.~(\ref{eq:transABSfinite}) takes into account the finite width of the N region. This term becomes only important in long junctions, $\Delta\gg\hbar v_F/L_N$, where it causes multiple subbands to appear in the Andreev spectrum. These additional bound states in long junctions cannot be captured with a $\delta$-model, which always yields two ABS per edge. Still for short junctions, $\Delta\ll\hbar v_F/L_N$, Eq.~(\ref{eq:transABSfinite}) tends to Eq.~(\ref{eq:transABSdelta}).

\subsection{Continuum states}\label{App:Continuum}
In the previous section, we have determined the discrete Andreev spectrum. If we now turn to the continuous spectrum with $\epsilon_s>\Delta$, we have to modify the ansatz~(\ref{eq:ABSansatz}) to account for propagating wave functions. An incident electron-like quasiparticle propagating from $x\to-\infty$ to $x\to\infty$ with $\epsilon_s>\Delta$ can be described by the ansatz
\begin{equation}\label{eq:conansatz}
\psi_{s,\sigma}(x)=\frac{1}{\sqrt{L_{tot}}}\left\{\begin{array}{l}
\left(\begin{array}{l}u_s\\v_s\end{array}\right)\e^{\i(k_F+q^{(s)}_\epsilon)x}+r_{eh}^{s,\sigma}\left(\begin{array}{l}v_s\\u_s\end{array}\right)\e^{\i(k_F-q^{(s)}_\epsilon)x},\; x<0,\\
\quad\\
\quad\\
e_{s,\sigma}\left(\begin{array}{l}1\\0\end{array}\right)\e^{\i(k_N+k^{(s)}_\epsilon)x}+h_{s,\sigma}\left(\begin{array}{l}0\\1\end{array}\right)\e^{\i(k_N-k^{(s)}_\epsilon)x},\; 0<x<L_N,
\quad\\
\quad\\
t_{ee}^{s,\sigma}\left(\begin{array}{l}u_s\\v_s\e^{-\i\varphi_\sigma}\end{array}\right)\e^{\i(k_F+q^{(s)}_\epsilon)x},\; x>L_N,
\end{array}\right.
\end{equation}
if a finite N region is considered. The ansatz for a $\delta$-junction is again similar to Eq.~(\ref{eq:conansatz}), but with $\psi_{s,\sigma}(x<0)$ given by the first line of Eq.~(\ref{eq:conansatz}) and $\psi_{s,\sigma}(x>0)$ given by the third line of Eq.~(\ref{eq:conansatz}). In order for the quasiparticle to be a right mover, $s\sigma=1$ in Eq.~(\ref{eq:conansatz}), that is, $s=\uparrow$ at the top edge and $s=\downarrow$ at the bottom edge. Corresponding equations can be set up for hole-like quasiparticles propagating to the right as well as for quasiparticles propagating to the left. In Eq.~(\ref{eq:conansatz}), $u_s^2=(1+\sqrt{\epsilon_s^2-\Delta^2}/\epsilon_s)/2=1-v_s^2$, $q^{(s)}_\epsilon=\sqrt{\epsilon_s^2-\Delta^2}/(\hbar v_F)$, and all other quantities are the same as defined in Eq.~(\ref{eq:ABSansatz})~\footnote{Note that no additional normalization factors are needed in Eq.~(\ref{eq:conansatz}) because each individual state in the S regions carries the same absolute value of quasiparticle current.}.

By invoking the boundary conditions~(\ref{eq:transABSdelta}) or~(\ref{eq:transABSfinite}), one can then obtain the reflection and transmission coefficients $r_{eh}^{s,\sigma}$ and $t_{ee}^{s,\sigma}$ in Eq.~(\ref{eq:conansatz}). For hole-like incident quasiparticles, one can likewise obtain $r_{he}^{s,\sigma}$ and $t_{hh}^{s,\sigma}$. These states then allow us to set up the $S$ matrix as
\begin{equation}\label{eq:Smat}
S_{SNS}=\left(\begin{array}{cc} S_{SNS}^t & 0 \\ 0 & S_{SNS}^b \end{array}\right)
\end{equation}
with
\begin{equation}\label{eq:Smat_t}
S_{SNS}^t=\left(\begin{array}{cccc} t_{ee}^{\uparrow,t} & 0 & 0 & r_{he}^{\uparrow,t} \\ 0 & t_{hh}^{\downarrow,t} & r_{eh}^{\downarrow,t} & 0 \\ 0 & r_{he}^{\downarrow,t} & t_{ee}^{\downarrow,t} & 0 \\ r_{eh}^{\uparrow,t} & 0 & 0 & t_{hh}^{\uparrow,t} \end{array}\right)
\end{equation}
and
\begin{equation}\label{eq:Smat_b}
S_{SNS}^b=\left(\begin{array}{cccc} t_{ee}^{\downarrow,b} & 0 & 0 & r_{he}^{\downarrow,b} \\ 0 & t_{hh}^{\uparrow,b} & r_{eh}^{\uparrow,b} & 0 \\ 0 & r_{he}^{\uparrow,b} & t_{ee}^{\uparrow,b} & 0 \\ r_{eh}^{\downarrow,b} & 0 & 0 & t_{hh}^{\downarrow,b} \end{array}\right).
\end{equation}

The scattering matrix $S^\sigma_{SNS}(\epsilon,\varphi)$ allows us then to compute the ($\varphi$-dependent part of the) continuum DOS of the Josephson junction as
\begin{equation}\label{eq:DOSphase}
\rho^\sigma_0(\epsilon,\varphi)=\frac{1}{2\pi\i}\frac{\partial}{\partial\epsilon}\ln\left[\det\left(S^\sigma_{SNS}\right)\right].
\end{equation}
Since $S_{SNS}$ is block-diagonal in the top/bottom-edge degree of freedom, $\mathrm{det}\left(S_{SNS}\right)=\mathrm{det}\left(S_{SNS}^t\right)\mathrm{det}\left(S_{SNS}^b\right)$ can be decomposed in a contribution from the top edge ($\sigma=1$) and one from the bottom edge ($\sigma=-1$). Then, we obtain
\begin{equation}\label{eq:DetSmat_t_delta}
\mathrm{det}\left(S_{SNS}^t\right)=\frac{\epsilon_\uparrow^2-\Delta^2\cos^2(\varphi_t/2)}{\epsilon_\uparrow^2\cos\varphi_t-\Delta^2\cos^2(\varphi_t/2)+\i\epsilon_\uparrow\sqrt{\epsilon_\uparrow^2-\Delta^2}\sin\varphi_t}\frac{\epsilon_\downarrow^2-\Delta^2\cos^2(\varphi_t/2)}{\epsilon_\downarrow^2\cos\varphi_t-\Delta^2\cos^2(\varphi_t/2)-\i\epsilon_\downarrow\sqrt{\epsilon_\downarrow^2-\Delta^2}\sin\varphi_t}
\end{equation}
for the top edge of a $\delta$-barrier model if $\epsilon_{\uparrow/\downarrow}>\Delta$. Note that because spin $s=\uparrow/\downarrow$ is a good quantum number, Eq.~(\ref{eq:DetSmat_t_delta}) also factorizes into separate contributions from $s=\uparrow/\downarrow$. The contribution from the bottom edge is also given by Eq.~(\ref{eq:DetSmat_t_delta}), but with the substitution $B\to-B$, that is, $\varphi_t\to\varphi_b$, $\epsilon_\uparrow\to\epsilon_\downarrow$, and $\epsilon_\downarrow\to\epsilon_\uparrow$. Inserting Eq.~(\ref{eq:DetSmat_t_delta}) into Eq.~(\ref{eq:DOSphase}) then yields Eq.~(\ref{eq:delta_con}) in the main text.

If we consider a finite N region, the above discussion still applies, but now
\begin{eqnarray}\label{eq:DetSmat_t_finite}
\mathrm{det}\left(S_{SNS}^t\right)=\frac{\e^{-2\i q^{(\uparrow)}_\epsilon L_N}\left[\epsilon_\uparrow^2-\Delta^2\cos^2\left(\frac{\varphi_t-2k^{(\uparrow)}_\epsilon L_N}{2}\right)\right]}{\epsilon_\uparrow^2\cos\left(\varphi_t-2k^{(\uparrow)}_\epsilon L_N\right)-\Delta^2\cos^2\left(\frac{\varphi_t-2k^{(\uparrow)}_\epsilon L_N}{2}\right)+\i\epsilon_\uparrow\sqrt{\epsilon_\uparrow^2-\Delta^2}\sin\left(\varphi_t-2k^{(\uparrow)}_\epsilon L_N\right)}\nonumber\\
\times\frac{\e^{-2\i q^{(\downarrow)}_\epsilon L_N}\left[\epsilon_\downarrow^2-\Delta^2\cos^2\left(\frac{\varphi_t+2k^{(\downarrow)}_\epsilon L_N}{2}\right)\right]}{\epsilon_\downarrow^2\cos\left(\varphi_t+2k^{(\downarrow)}_\epsilon L_N\right)-\Delta^2\cos^2\left(\frac{\varphi_t+2k^{(\downarrow)}_\epsilon L_N}{2}\right)-\i\epsilon_\downarrow\sqrt{\epsilon_\downarrow^2-\Delta^2}\sin\left(\varphi_t+2k^{(\downarrow)}_\epsilon L_N\right)}
\end{eqnarray}
for the top edge if $\epsilon_{\uparrow/\downarrow}>\Delta$. The contribution from the bottom edge can be obtained by replacing $\varphi_t\to\varphi_b$, $\epsilon_\uparrow\to\epsilon_\downarrow$, and $\epsilon_\downarrow\to\epsilon_\uparrow$.

\section{Free energy and thermodynamic observables}\label{App:FE}
From the ABS and continuum spectra of Eq.~(\ref{eq:BDG}) one can calculate the free energy $F_\sigma(\varphi,T)$ at a single edge $\sigma=t/b$. These free energies can in turn be used to obtain the Josephson current and other thermodynamic quantities of a given edge.

\subsection{No parity constraints}\label{App:FENoParity}
Without parity constraints, the free energy is---up to an additive $\varphi$-independent contribution---given by~\cite{Beenakker1992,Beenakker2013:PRL}
\begin{equation}\label{eq:FE}
F^\sigma_0(\varphi,T)=-k_BT\left\{\sum\limits_{n,\epsilon^\sigma_n\geq0}\ln\left[2\cosh\left(\frac{\epsilon^\sigma_n(\varphi)}{2k_BT}\right)\right]+\int\limits_0^\infty\d\epsilon\rho^\sigma_c(\epsilon,\varphi)\ln\left[2\cosh\left(\frac{\epsilon}{2k_BT}\right)\right]\right\},
\end{equation}
where $k_B$ is the Boltzmann constant and $T$ the temperature. The sum over $n$ describes the contribution from the ABS with discrete energies $\epsilon^\sigma_n(\varphi)$, while the integral describes the contribution from the continuum states with DOS $\rho^\sigma_c(\epsilon,\varphi)$. At this point, it is important to note that due to the lack of spin degeneracy the degeneracy factor is just $g=1$ in Eq.~(\ref{eq:FE})~\cite{Beenakker2013:PRL}. Whereas we can directly insert the Andreev spectrum $\epsilon^\sigma_n(\varphi)$ in Eq.~(\ref{eq:FE}), we have to compute the continuum DOS
\begin{equation}\label{eq:DOS}
\rho^\sigma_c(\epsilon,\varphi)=\rho^\sigma_0(\epsilon,\varphi)+\rho_S(\epsilon).
\end{equation}
The $\varphi$-dependent continuous spectrum of the Josephson junction is described by $\rho^\sigma_0(\epsilon,\varphi)$, which we compute from the scattering matrix via Eq.~(\ref{eq:DOSphase}). Following Ref.~\cite{Beenakker2013:PRL}, we also include in Eq.~(\ref{eq:DOS}) a $\varphi$-independent term originating from the superconducting electrodes,
\begin{equation}\label{eq:DOSSv2}
\rho_S(\epsilon)=\frac{2}{\pi E_S}\frac{|\epsilon|\Theta\left(\epsilon^2-\Delta^2\right)}{\sqrt{\epsilon^2-\Delta^2}},
\end{equation}
see also Eq.~(\ref{eq:DOSS}) in the main text. Here the energy scale $E_S=\hbar v_F/L_S$ is determined from the length $L_S$ of the superconducting electrodes. While $\rho_S(\epsilon)$ is independent of $\varphi$ and does not affect the Josephson current in the absence of parity constraints~\cite{Beenakker1992}, it can play an important role if the parity is kept constant.

\subsection{Parity constraints}\label{Sec:FEParity}
If the fermion parity $p=\pm$ is conserved at a given edge $\sigma=t/b$, the free energy acquires an additional contribution due to the parity constraint and reads~\cite{Ioselevich2011:PRL,Beenakker2013:PRL}
\begin{equation}\label{eq:FEp}
F^\sigma_p(\varphi,T)=F^\sigma_0(\varphi,T)-k_BT\ln\left\{\frac{1}{2}\left\{1+pP(\varphi)\left[\prod\limits_{n,\epsilon^\sigma_n\geq0}\tanh\left(\frac{\epsilon^\sigma_n(\varphi)}{2k_BT}\right)\right]\exp\left[\int\limits_0^\infty\d\epsilon\rho^\sigma_c(\epsilon,\varphi)\ln\left(\tanh\left(\frac{\epsilon}{2k_BT}\right)\right)\right]\right\}\right\}.
\end{equation}
Here $p=\pm1$ and the function $P(\varphi)$ describes the ground-state fermion parity as $\varphi$ is tuned, see also Sec.~\ref{Sec:DeltaParity} in the main text. From the form of Eq.~(\ref{eq:FEp}) it is clear that even $\varphi$-independent terms in $\rho^\sigma_c$ are important in determining the $\varphi$ dependence of $F^\sigma_p$. This is why, for example, the contribution from the superconducting electrodes can also affect other quantities such as the Josephson current. If we split the integral over $\rho^\sigma_c$ in Eq.~(\ref{eq:FEp}) into an integral over $\rho^\sigma_0$ and $\rho_S$, define
\begin{equation}\label{eq:JsDef}
J_S(T)=\int\limits_0^\infty\d\epsilon\rho_S(\epsilon)\ln\left[\tanh\left(\frac{\epsilon}{2k_BT}\right)\right]=-\frac{2}{\pi k_BTE_S}\int\limits_\Delta^\infty\d\epsilon\;\frac{\sqrt{\epsilon^2-\Delta^2}}{\sinh\left(\epsilon/k_BT\right)},
\end{equation}
and insert the ABS for $\epsilon^\sigma_n$, we arrive at Eq.~(\ref{eq:deltaFParity}) in the main text.

\begin{figure}[ht]
\centering
\includegraphics*[width=13.0cm]{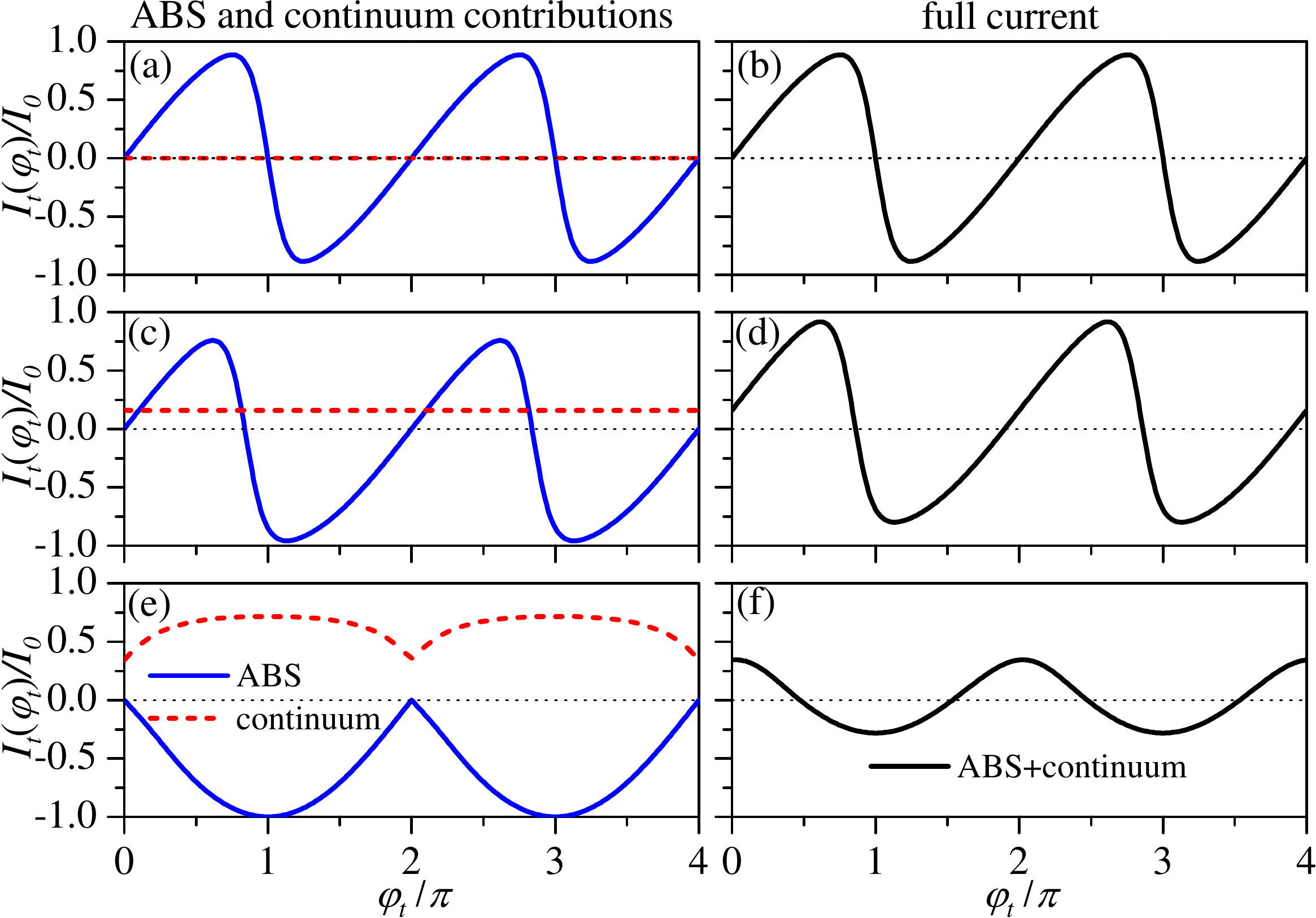}
\caption{(Color online) Different contributions and full Josephson current of the top edge as a function of $\varphi\equiv\varphi_t$ for $k_BT=0.1\Delta$: (a,b) $v_Fp_S=0$, (c,d) $v_Fp_S=0.5\Delta$, (e,f) $v_Fp_S=2.5\Delta$. Here we have used the $\delta$-barrier model and $I_0=e\Delta/2\hbar$.}\label{fig:deltaJC}
\end{figure}

\begin{figure}[ht]
\centering
\includegraphics*[width=13.0cm]{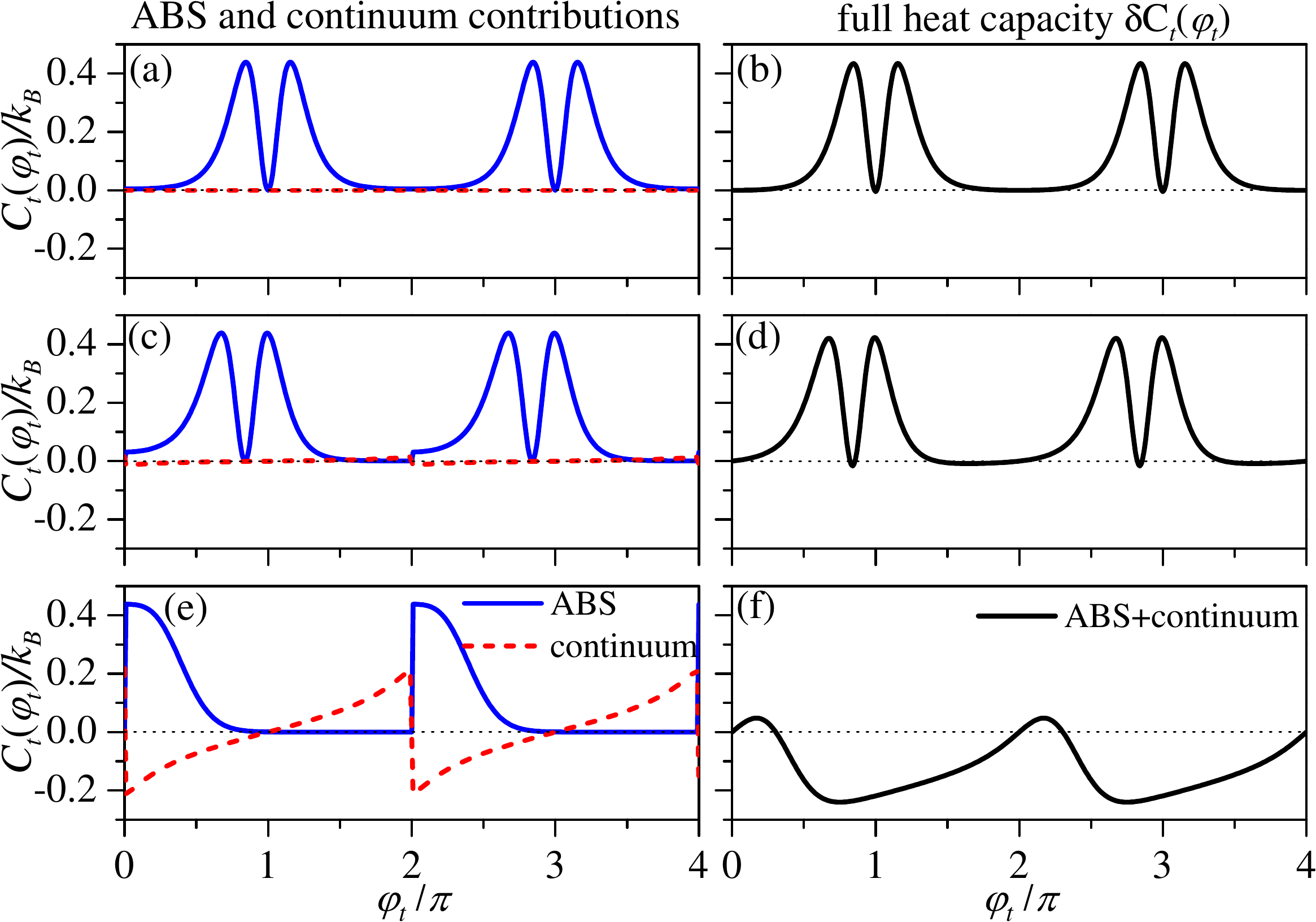}
\caption{(Color online) Different contributions and full heat capacity of the top edge as a function of $\varphi\equiv\varphi_t$ for $k_BT=0.1\Delta$: (a,b) $v_Fp_S=0$, (c,d) $v_Fp_S=0.5\Delta$, (e,f) $v_Fp_S=2.5\Delta$. Here we have used the $\delta$-barrier model. In panels~(a,c,e), only the $\varphi$-dependent parts of $C_{ABS}^t(\varphi,T)$ and $C_c^t(\varphi,T)$ are plotted. In panels~(b,d,f), $\Delta C_t(\varphi)=C_t(\varphi)-C_t(0)$.}\label{fig:deltaHC}
\end{figure}

\subsection{Josephson current and entropy at a single edge}\label{App:FECurrentEntropy}
Having determined $F_\sigma$, given either by Eq.~(\ref{eq:FE}) or by Eq.~(\ref{eq:FEp}), we can then calculate the Josephson current via
\begin{equation}\label{eq:JoCurv2}
I_\sigma(\varphi,T)=\frac{2e}{\hbar}\frac{\partial F_\sigma(\varphi,T)}{\partial\varphi},
\end{equation}
where $e$ is the elementary charge, as well as the entropy via
\begin{equation}\label{eq:Entropyv2}
S_\sigma(\varphi,T)=-\frac{\partial F_\sigma(\varphi,T)}{\partial T}.
\end{equation}
Due to the Maxwell relation
\begin{equation}\label{eq:MR}
-\frac{\partial^2F_\sigma}{\partial T\partial\varphi}=\frac{\partial S_\sigma}{\partial\varphi}=-\frac{\hbar}{2e}\frac{\partial I_\sigma}{\partial T},
\end{equation}
the phase dependence of the entropy can be determined from the Josephson current as
\begin{equation}\label{eq:MRI}
S_\sigma(\varphi)-S_\sigma(0)=-\frac{\hbar}{2e}\int\limits_0^\varphi\d\varphi'\;\frac{\partial I_\sigma(\varphi')}{\partial T}.
\end{equation}
Thus, one can use the temperature dependence of $I_\sigma$ to obtain the $\varphi$ dependence of $S_\sigma$. In addition, the heat capacity
\begin{equation}\label{eq:HeatCapv2}
C_\sigma(\varphi,T)=T\frac{\partial S_\sigma(\varphi,T)}{\partial T}
\end{equation}
provides a quantity that is directly accessible experimentally.

\section{Decomposing the Josephson current and heat capacity into contributions from the Andreev bound states and the continuum}\label{App:Decomp}
As mentioned in the main text, one can separate the contributions of the ABS and the continuum states to the free energy $F_0^t$ from each other if there are no parity constraints. Consequently, one can also split the Josephson current, the entropy and the heat capacity in two such separate contributions [see Eqs.~(\ref{eq:delta_JCABS}), (\ref{eq:delta_JCcon}), (\ref{eq:delta_CABS}), and (\ref{eq:delta_Ccon}) in the main text].

For illustration, Fig.~\ref{fig:deltaJC} shows the separate contributions $I_{ABS}^t$ and $I_c^t$ as well as the full Josephson current $I_t=I_{ABS}^t+I_c^t$ for different Doppler shifts $v_Fp_S$. Whereas in Figs.~\ref{fig:deltaJC}(a,b) with $p_S=0$ the total current is given by $I_t=I_{ABS}^t$, the situation changes for $p_S\neq0$: For finite $p_S$, some weight of the DOS is transferred from the ABS to the continuum states~\cite{Zhao2003:PRL,Zhao2004:PRB}, and a finite $I_c^t$ develops. If $0<|v_Fp_S|<2\Delta$, the effect of the continuum states is to mainly shift the total current $I_t$ compared to $I_{ABS}^t$ as shown in Figs.~\ref{fig:deltaJC}(c,d). As $p_S$ is increased further, such that $|v_Fp_S|>2\Delta$, $I_{ABS}^t$ becomes unidirectional [see Fig.~\ref{fig:deltaJC}(e)], in the sense that $I_{ABS}^t$ flows in the same direction for any phase difference $\varphi_t$. This is similar to Ref.~\cite{Tkachov2019:PRB}, where the unidirectional, chiral current carried by the ABS has been shown as a hallmark of a $4\pi$-periodic current-phase relation. The total Josephson current, on the other hand, is a smooth function of $\varphi_t$ and no longer unidirectional due to the contribution from the continuum states. These total Josephson currents are the ones shown in Fig.~\ref{fig:DeltaCHC}(a) of the main text.

Similarly, the total heat capacity of the top edge computed via Eqs.~(\ref{eq:Entropy}) and~(\ref{eq:HeatCap}) can also be split into contributions $C_{ABS}^t$ from the ABS and contributions $C_c^t$ from the continuum [see Eqs.~(\ref{eq:delta_CABS}) and (\ref{eq:delta_Ccon}) in the main text]. Figure~\ref{fig:deltaHC} shows $C_{ABS}^t$ and $C_c^t$ as well as the full heat capacity $C_t=C_{ABS}^t+C_c^t$ for different values of $p_S$. Assuming $\Delta(T)\approx\Delta(T=0)$, we present $C_{ABS}^t$ and $C_c^t$ up to a $\varphi_t$-independent constant, while the total heat capacity of the top edge $C_t$ is measured with respect to its value at $\varphi_t=0$. For $v_Fp_S=0$, the heat capacity in $\delta$-junctions is again exclusively due to the ABS as shown in Figs.~\ref{fig:deltaHC}(a,b). Importantly, one can see that around the zero-energy crossing of the ABS, $\varphi_t=\pi$, a sharp double-peak feature given by Eq.~(\ref{eq:delta_CABS}) develops.

For finite $p_S$, the continuum contribution plays a crucial role: As one can discern from Figs.~\ref{fig:deltaHC}(c,e), $C_{ABS}^t$ is discontinuous due to the discontinuities of the ABS spectrum~(\ref{eq:delta_ABS}) if $\varphi_t$ is an integer multiple of $2\pi$. These apparent discontinuities are artificial, however, since the ABS merge into the continuum at these values of $\varphi_t$. Indeed, including the continuum contribution~(\ref{eq:delta_Ccon}) lifts these discontinuities in the total heat capacity $C_t$ as demonstrated by Figs.~\ref{fig:deltaHC}(d,f). If $0<|v_Fp_S|<2\Delta$, $C_t$ is qualitatively very similar to the case of $p_S=0$ with a sharp double-peak feature around the phase $\varphi_t$, where the ABS cross at zero energy. If $|v_Fp_S|>2\Delta$, the ABS no longer cross at zero energy as shown in Fig.~\ref{fig:ABS}(c) and consequently $C_t$ no longer exhibits the sharp double-peak feature characteristic of the zero-energy crossing [see Fig.~\ref{fig:deltaHC}(f)].

\begin{figure}[ht]
\centering
\includegraphics*[width=13cm]{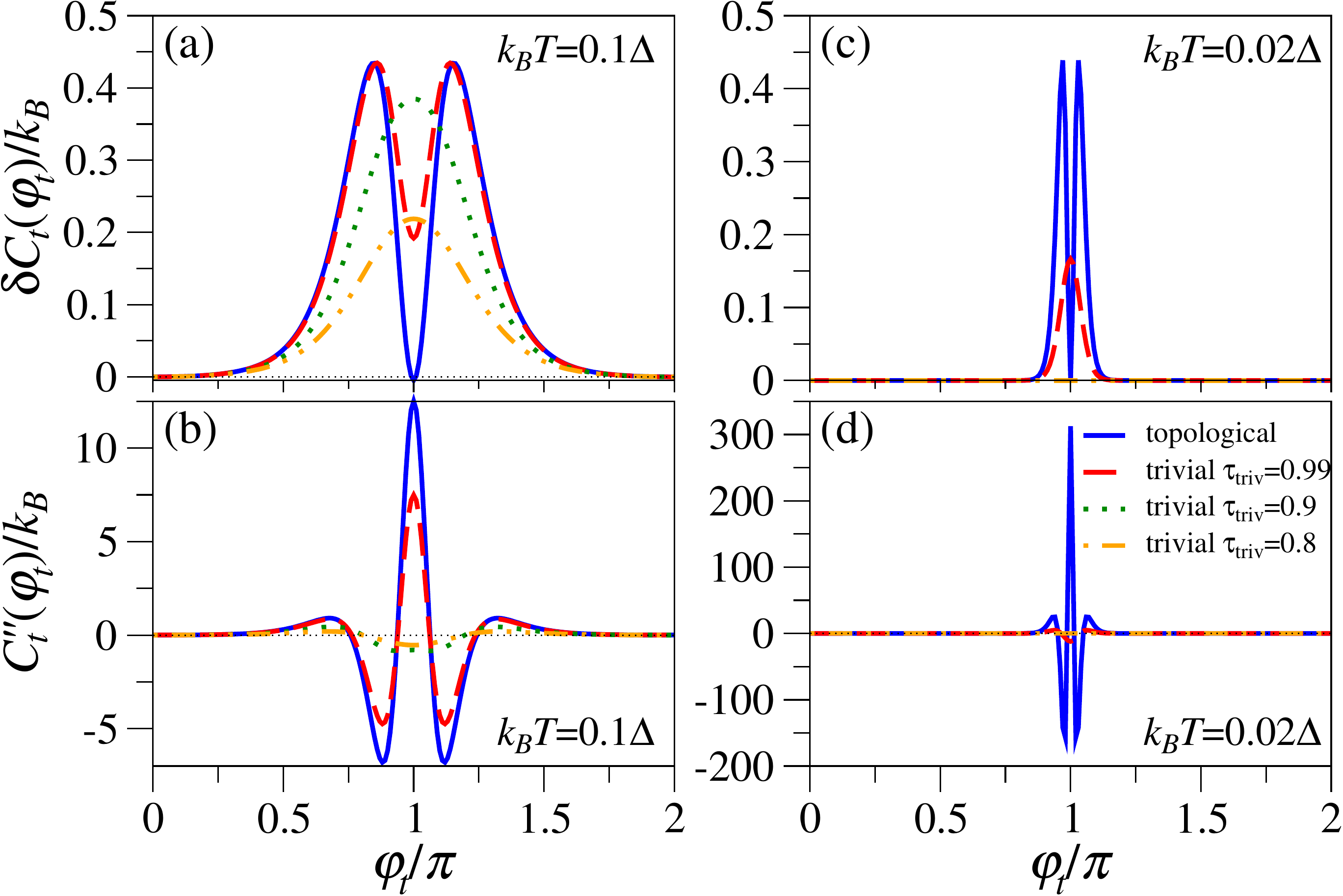}
\caption{(Color online) Phase dependence of (a,c) the relative heat capacity $\delta C^t(\varphi_t)=C^t(\varphi_t)-C^t(0)$ and (b,d) $C^{''}_t(\varphi_t)=\d^2C^t(\varphi_t)/\d\varphi_t^2$ for the top edge of a topological Josephson junction and for a trivial Josephson junction with different transmission probabilities $\tau_\mathrm{triv}$. Here $p_S=0$ and no parity constraints have been imposed on the topological junction. The temperature has been chosen as $k_BT=0.1\Delta$ in panels~(a,b) and as $k_BT=0.02\Delta$ in panels~(c,d).}\label{fig:TopVsTriv_Phi}
\end{figure}

\section{Comparison between the heat capacities of a topological and a trivial Josephson junction}\label{App:TopVsTriv}
In the absence of parity constraints, the heat capacity $C^t(\varphi_t)$ of a topological Josephson junction exhibits pronounced double-peak features as a function of the phase bias $\varphi_t$ for $k_BT\ll\Delta$, providing a signature of the protected zero-energy crossing. Here we show that this is in contrast to the behavior of the heat capacity in a trivial Josephson junction. For simplicity, we compare the topological Josephson junction discussed in the main text without parity constraints and for $p_S=0$ to a one-dimensional (1D) trivial Josephson junction.

We consider a short 1D trivial Josephson junction based on a quadratic Hamiltonian. Such a junction exhibits a pair of ABS per spin with energies~\cite{Beenakker1992}
\begin{equation}\label{eq:delta_trivABS}
\epsilon^{\pm}_\mathrm{triv}(\varphi_t)=\pm\Delta\sqrt{1-\tau_\mathrm{triv}\sin^2\left(\frac{\varphi_t}{2}\right)},
\end{equation}
where $0\leq\tau_\mathrm{triv}\leq1$ describes the junction transparency (modulated, for example, by a potential difference $V_0$ between the N and S regions). Since we are studying the limit of short junctions, we assume that the $\varphi_t$ dependence of the free energy originates solely from the ABS. Then, the heat capacity (per spin) of the trivial junction is also described by Eq.~(\ref{eq:delta_CABS}) in the main text, with $\epsilon^t_\uparrow(\varphi_t)$ replaced by $\epsilon^+_\mathrm{triv}(\varphi_t)$.

For a perfectly transparent junction, $\tau_\mathrm{triv}=1$ and Eq.~(\ref{eq:delta_trivABS}) reduces to $\epsilon^{\pm}_\mathrm{triv}(\varphi_t)=\pm\Delta\cos(\varphi_t/2)$, similar to the case of the ABS in a short topological junction for $p_S=0$, see Eq.~(\ref{eq:delta_ABS}) in the main text. In contrast to a topological junction, the zero-energy crossing at $\varphi=\pi$ is not protected in a trivial junction and is removed if $\tau_\mathrm{triv}\neq1$. This results in a gap of $\delta\epsilon=2\Delta\sqrt{1-\tau_\mathrm{triv}}$.

In Fig.~\ref{fig:TopVsTriv_Phi}, we show $C^t(\varphi_t)$ (measured with respect to $\varphi_t=0$) and its second derivative, $C^{''}_t(\varphi_t)=\d^2C^t(\varphi_t)/\d\varphi_t^2$, for a topological Josephson junction as well as for a trivial junction with different values of $\tau_\mathrm{triv}$. As long as $k_BT\ll\delta\epsilon$, the splitting $\delta\epsilon$ can be thermally resolved and a peak develops at $\varphi_t=\pi$ for a trivial Josephson junction as illustrated by Fig.~\ref{fig:TopVsTriv_Phi}(a) for $\tau_\mathrm{triv}=0.8$ and $\tau_\mathrm{triv}=0.9$. If $k_BT$ is too large to clearly resolve $\delta\epsilon$, as is the case in Figs.~\ref{fig:TopVsTriv_Phi}(a,b) for $\tau_\mathrm{triv}=0.99$, $C^t(\varphi_t)$ resembles that of a topological junction. If the temperature is decreased, however, the double-peak feature around $\varphi_t=\pi$ merges into a single peak at $\varphi_t=\pi$ as can be seen by comparing Figs.~\ref{fig:TopVsTriv_Phi}(a) and~(c). In a topological junction, on the other hand, the double-peak feature with a minimum at $\varphi_t=\pi$ remains as $T$ is decreased. Hence, probing the $T$ dependence of $C^t(\varphi_t)$ around $\varphi_t=\pi$ allows to distinguish between topological and trivial Josephson junctions: $C^t(\varphi_t)$ always exhibits a minimum at $\varphi_t=\pi$ in a topological junction, while a peak develops at $\varphi_t=\pi$ in a trivial junction as $T$ is decreased.

\end{widetext}

\bibliography{BibTopInsAndTopSup}

\end{document}